\documentclass[preprint]{aastex}
\usepackage{natbib}
\usepackage{multirow}
\usepackage{hyperref}
\usepackage{enumitem}
\usepackage{unites2e}
\usepackage{lineno}
\usepackage{textcomp}
\usepackage{xcolor}
\newcommand{\hours}{\ensuremath{^\mathrm{h}}}
\newcommand{\minutes}{\ensuremath{^\mathrm{m}}}
\newcommand{\seconds}{\ensuremath{^\mathrm{s}}}

\shorttitle{Observations of Cas A with VERITAS}
\shortauthors{Kumar et al.}

\begin{document}


\title{Evidence for proton acceleration up to TeV energies based on VERITAS and \textit{Fermi}-LAT observations of the Cas~A SNR }

\author{
A.~U.~Abeysekara\altaffilmark{1},
A.~Archer\altaffilmark{2},
W.~Benbow\altaffilmark{3},
R.~Bird\altaffilmark{4},
R.~Brose\altaffilmark{5,6},
M.~Buchovecky\altaffilmark{4},
J.~H.~Buckley\altaffilmark{7},
A.~J.~Chromey\altaffilmark{8},
W.~Cui\altaffilmark{9,10},
M.~K.~Daniel\altaffilmark{3},
S.~Das\altaffilmark{11},
V.~V.~Dwarkadas\altaffilmark{12},
A.~Falcone\altaffilmark{13},
Q.~Feng\altaffilmark{14},
J.~P.~Finley\altaffilmark{9},
L.~Fortson\altaffilmark{15},
A.~Gent\altaffilmark{16},
G.~H.~Gillanders\altaffilmark{17},
C.~Giuri\altaffilmark{6},
O.~Gueta\altaffilmark{6},
D.~Hanna\altaffilmark{11},
T.~Hassan\altaffilmark{6},
O.~Hervet\altaffilmark{18},
J.~Holder\altaffilmark{19},
G.~Hughes\altaffilmark{3},
T.~B.~Humensky\altaffilmark{14},
P.~Kaaret\altaffilmark{20},
P.~Kar\altaffilmark{1},
N.~Kelley-Hoskins\altaffilmark{6},
M.~Kertzman\altaffilmark{2},
D.~Kieda\altaffilmark{1},
M.~Krause\altaffilmark{6},
F.~Krennrich\altaffilmark{8},
S.~Kumar\altaffilmark{11},
M.~J.~Lang\altaffilmark{17},
G.~Maier\altaffilmark{6},
P.~Moriarty\altaffilmark{17},
R.~Mukherjee\altaffilmark{21},
M.~Nievas-Rosillo\altaffilmark{6},
S.~O'Brien\altaffilmark{11},
R.~A.~Ong\altaffilmark{4},
N.~Park\altaffilmark{22},
A.~Petrashyk\altaffilmark{14},
K.~Pfrang\altaffilmark{6},
M.~Pohl\altaffilmark{5,6},
E.~Pueschel\altaffilmark{6},
J.~Quinn\altaffilmark{23},
K.~Ragan\altaffilmark{11},
P.~T.~Reynolds\altaffilmark{24},
G.~T.~Richards\altaffilmark{19},
E.~Roache\altaffilmark{3},
I.~Sadeh\altaffilmark{6},
M.~Santander\altaffilmark{25},
G.~H.~Sembroski\altaffilmark{9},
K.~Shahinyan\altaffilmark{15},
I.~Sushch\altaffilmark{5},
A.~Weinstein\altaffilmark{8},
P.~Wilcox\altaffilmark{20},
A.~Wilhelm\altaffilmark{5,6},
D.~A.~Williams\altaffilmark{18},
T.~J~Williamson\altaffilmark{19},
B.~Zitzer\altaffilmark{11}
A.~Ghiotto\altaffilmark{14}
}

\altaffiltext{1}{Department of Physics and Astronomy, University of Utah, Salt Lake City, UT 84112, USA}
\altaffiltext{2}{Department of Physics and Astronomy, DePauw University, Greencastle, IN 46135-0037, USA}
\altaffiltext{3}{Center for Astrophysics $|$ Harvard \& Smithsonian, Cambridge, MA 02138, USA}
\altaffiltext{4}{Department of Physics and Astronomy, University of California, Los Angeles, CA 90095, USA}
\altaffiltext{5}{Institute of Physics and Astronomy, University of Potsdam, 14476 Potsdam-Golm, Germany}
\altaffiltext{6}{DESY, Platanenallee 6, 15738 Zeuthen, Germany}
\altaffiltext{7}{Department of Physics, Washington University, St. Louis, MO 63130, USA}
\altaffiltext{8}{Department of Physics and Astronomy, Iowa State University, Ames, IA 50011, USA}
\altaffiltext{9}{Department of Physics and Astronomy, Purdue University, West Lafayette, IN 47907, USA}
\altaffiltext{10}{Department of Physics and Center for Astrophysics, Tsinghua University, Beijing 100084, China.}
\altaffiltext{11}{Physics Department, McGill University, Montreal, QC H3A 2T8, Canada}
\altaffiltext{12}{Department of Astronomy and Astrophysics, University of Chicago, Chicago, IL, 60637}
\altaffiltext{13}{Department of Astronomy and Astrophysics, 525 Davey Lab, Pennsylvania State University, University Park, PA 16802, USA}
\altaffiltext{14}{Physics Department, Columbia University, New York, NY 10027, USA}
\altaffiltext{15}{School of Physics and Astronomy, University of Minnesota, Minneapolis, MN 55455, USA}
\altaffiltext{16}{School of Physics and Center for Relativistic Astrophysics, Georgia Institute of Technology, 837 State Street NW, Atlanta, GA 30332-0430}
\altaffiltext{17}{School of Physics, National University of Ireland Galway, University Road, Galway, Ireland}
\altaffiltext{18}{Santa Cruz Institute for Particle Physics and Department of Physics, University of California, Santa Cruz, CA 95064, USA}
\altaffiltext{19}{Department of Physics and Astronomy and the Bartol Research Institute, University of Delaware, Newark, DE 19716, USA}
\altaffiltext{20}{Department of Physics and Astronomy, University of Iowa, Van Allen Hall, Iowa City, IA 52242, USA}
\altaffiltext{21}{Department of Physics and Astronomy, Barnard College, Columbia University, NY 10027, USA}
\altaffiltext{22}{WIPAC and Department of Physics, University of Wisconsin-Madison, Madison WI, USA}
\altaffiltext{23}{School of Physics, University College Dublin, Belfield, Dublin 4, Ireland}
\altaffiltext{24}{Department of Physical Sciences, Cork Institute of Technology, Bishopstown, Cork, Ireland}
\altaffiltext{25}{Department of Physics and Astronomy, University of Alabama, Tuscaloosa, AL 35487, USA}



\begin{abstract}
We present a study of $\gamma$-ray emission from the core-collapse supernova remnant Cas~A in the energy range from $0.1 \U{GeV}$ to $10 \U{TeV}$. We used 65 hours of VERITAS data to cover 200 GeV - 10 TeV, and 10.8 years of \textit{Fermi}-LAT data to cover 0.1-500 GeV. The spectral analysis of \textit{Fermi}-LAT data shows a significant spectral curvature around $1.3 \pm 0.4_{stat} \U{GeV}$ that is consistent with the expected spectrum from pion decay. Above this energy, the joint spectrum from \textit{Fermi}-LAT and VERITAS deviates significantly from a simple power-law, and is best described by a power-law with spectral index of $2.17\pm 0.02_{stat}$ with a cut-off energy of $2.3 \pm 0.5_{stat} \U{TeV}$. These results, along with radio, X-ray and $\gamma$-ray data, are interpreted in the context of leptonic and hadronic models. Assuming a one-zone model, we exclude a purely leptonic scenario and conclude that proton acceleration up to at least 6 TeV is required to explain the observed $\gamma$-ray spectrum. From modeling of the entire multi-wavelength spectrum, a minimum magnetic field inside the remnant of $B_{\mathrm{min}}\approx150\,\mathrm{\mu G}$ is deduced.
\end{abstract}


\section{INTRODUCTION}
Supernova remnants (SNRs) are considered to be the most promising sites for the acceleration of Galactic cosmic rays up to PeV ($10^{15} \U{eV}$) energies, since they can provide sufficient energy to maintain the cosmic-ray energy flux in our Galaxy \citep{Baade1934, Ginzburg_1966}. Additional support for this idea is given by the fact that diffusive shock acceleration (DSA) mechanism \citep{Krymskii1977,Axford1977,Bell1978I,Bell1978II,Blandford1978}, believed to occur at SNR shocks, predicts a particle spectrum in rough agreement with the observed cosmic ray spectrum corrected for propagation effects. As cosmic rays are charged particles due to which their path is deflected by the Galactic magnetic field, direct measurements cannot determine their point of origin; however, $\gamma$-rays, a neutral by-product of the interaction of cosmic rays with the medium around the source region, travel directly from their source of origin to a detector on Earth and thus provide a powerful tool to probe the origin of Galactic cosmic rays \citep{DEGRANGE2015}. 

Cassiopeia A (Cas~A) is the remnant of a core-collapse Type IIb supernova explosion \citep{Krause2008} that occurred in our Galaxy approximately $350$ years ago \citep{Fesen2006}. The progenitor of Cas~A is believed to have been a red supergiant, which lost most of its hydrogen envelope through strong stellar winds before the supernova occurred \citep{Chevalier2003}. Based on the proper motion of optical filaments, the distance to this SNR is estimated to be $3.4_{-0.1}^{+0.3}\U{kpc}$ \citep{Reed1995}, which leads to a physical size of the remnant of $\sim 5\U{pc}$ in diameter. Of the few historic Galactic SNRs, it has been observed extensively over a broad spectral range from radio through X-ray, and up to $\gamma$-ray wavelengths. 

The bright radio emission forming a circle of radius $\approx 1.7 \arcmin$ marked the location of ejecta interacting with the reverse shock in Cas~A \citep{Bell1975, Baars1977, Braun1987,Kassim1995}. Moreover, a fainter radio emission extended up to a radius of $\approx 2.5 \arcmin$ is also observed \citep{Delaney2014}. This radio emission has been interpreted as synchrotron radiation emitted by electrons moving in a magnetic field. Synchrotron emission from Cas~A is also detected in the near-infrared (IR) at $2.2 \U{\mu m}$ (K-band) \citep{Gerardy2001,Rho2003, Jones2003}. The dominant feature at near-IR wavelengths is diffuse emission that forms a complete ring and correlates well with the radio emission. Broadband spectral measurements from radio up to IR show a significant curvature, suggesting that the shock dynamics might have been modified by the back reaction of accelerated cosmic rays \citep{Rho2003}. 


The \textit{Chandra} X-ray Observatory has detected non-thermal X-ray emission in the shape of narrow rim at the forward shock at an energy of $4-6 \U{keV}$. This rim marked the boundary of the X-ray remnant; implying a size of $2.5 \arcmin \pm 0.2 \arcmin$ in radius \citep{Gotthelf2001}. The X-ray emission is interpreted as synchrotron radiation emitted by electrons accelerated to a maximum energy of $\sim 40-60 \U{TeV}$ at the forward shock \citep{Gotthelf2001, Vink_MF}. Along with the firm detection of non-thermal X-ray emission in the forward shock region, strong evidence was also found for non-thermal X-ray emission from the reverse shock region, primarily the western part \citep{Uchiyama2008, RSxray}. Recently, X-ray observations from \textit{NuSTAR} resolved the remnant above $15 \U{keV}$, finding that the emission is produced by knots located in the interior of the remnant \citep{Grefenstette2014}. Ten years of \textit{INTEGRAL} data published by  \citet{Wang2016} also showed non-thermal X-ray continuum emission, which can be fitted by a smooth power-law with no cut-off up to $220 \U{keV}$. Besides the non-thermal X-ray emission, there is also a strong thermal X-ray component, dominated mainly by line emission from the plasma of the shocked metal-rich ejecta \citep{Holt1994, Hwang2004}. Diffuse thermal emission has been studied by \cite{Lee2014} using  \textit{Chandra} X-ray observations. They determine that the thermal emission arises from the shocked circumstellar gas and is consistent with the model of an SNR interacting with a red supergiant wind. 

While non-thermal X-ray observations constrain the properties of the relativistic electron population, $\gamma$-ray observations can play an important role in determining the efficiency of proton acceleration at the shocks. High-energy protons produce $\gamma$-rays through the decay of neutral pions generated in collisions with ambient target material. However, $\gamma$-rays can also be produced by energetic electrons, through inverse-Compton (IC) scattering or non-thermal bremsstrahlung (NTB), which creates an ambiguity regarding the nature of the particle population producing the $\gamma$-ray emission. Precise measurements of the $\gamma$-ray emission spectrum, coupled with broadband spectral modeling, may allow us to resolve this ambiguity. Observations of two SNRs, IC443 and W44, by the Large Area Telescope (LAT) on board the \textit{Fermi} Gamma-ray Space Telescope, have reported the characteristic pion-decay signature of accelerated hadrons  in the $\gamma$-ray spectrum \citep{Ackermann2013}. 

The first detection of Cas~A as a $\gamma$-ray emitter in the MeV-GeV range was reported by \textit{Fermi}-LAT using one year of data \citep{Abdo2010}. Subsequently, with the data taken from 3.6 years of \textit{Fermi}-LAT observations, a detailed spectral analysis in the $0.1-100 \U{GeV}$ range was performed, showing a statistically significant break in the spectrum at $1.72_{-0.89}^{+1.35} \U{GeV}$ \citep{Yuan2013, Saha_2014}. Similar results were found from a recent analysis of $\sim 8$ years of \textit{Fermi}-LAT data by \citet{Ahnen2017}. 
At TeV energies, the first detection of Cas~A was made by the HEGRA stereoscopic Cherenkov telescope system \citep{Aharonian2001}. The differential photon spectrum measured between $1 \U{TeV}$ and $10\U{TeV}$ is consistent with a power-law (PL) with an index of $2.5 \pm 0.4_{stat} \pm 0.1_{sys}$ and the derived integral flux above $1 \U{TeV}$ is $(5.8 \pm 1.2_{stat} \pm 1.2_{sys}) \times 10^{-13} \U{cm^{-2}\: s^{-1}}$. These results were later confirmed by MAGIC \citep{Albert2007} and VERITAS \citep{Acciari2010}. 
Recently, a PL spectral index of $\Gamma = 2.8 \pm 0.1_{stat} \pm 0.2_{sys}$ was measured with an updated VERITAS data analysis \citep{KumarICRC2015} above $200 \U{GeV}$. This index is softer than the index of $2.2 \pm 0.1_{stat} \pm 0.1_{sys}$ measured by \citet{Yuan2013} above $2 \U{GeV}$, which indicates a spectral index change in the $\gamma$-ray spectrum around few hundred GeV. In 2017, the MAGIC collaboration showed that the PL distribution with a exponential cut-off is preferable over a single PL distribution with 4.6 standard deviation. They reported a spectral cut-off energy of $3.5_{-1.0}^{+1.6} \U{TeV}$ \citep{Ahnen2017}.Based on this result, \citet{Ahnen2017} suggest that Cas~A could not be a PeVatron at its present age. A caveat to this statement can be found in the work of \citet{2019ApJ...874...98Z} who note that a two-zone model for Cas A with specific assumptions may allow a proton cut off around 3 PeV.
 

Cas~A is assumed to be a point-like source for $\gamma$-ray instruments. This is because the size of remnant as measured in X-ray and radio ($\approx 150 \arcsec$ in radius) is comparable to the point spread function (PSF) of the $\gamma$-ray instruments. The  location of the peak of the $\gamma$-ray emission has been reported by various space-based and ground-based instruments. At $\U{GeV}$ energies, \citet{Yuan2013} reported the best-fit source position as right ascension (RA) $= (23\hours23\minutes24.7\seconds) \pm (0\hours0\minutes36.0\seconds)_{stat} \pm (0\hours0\minutes18.0\seconds)_{sys}$ and declination (Dec) $= (+58\arcdeg49\arcmin32.8\arcsec) \pm (0\arcdeg0\arcmin36.0\arcsec)_{stat} \pm (0\arcdeg0\arcmin18.0\arcsec)_{sys}$. In the $\U{TeV}$ range, VERITAS gives the centroid location as RA $= (23\hours23\minutes18.0\seconds)  \pm (0\hours0\minutes36.0\seconds)_{stat} \pm (0\hours1\minutes12.0\seconds)_{sys}$ and Dec $= (+58\arcdeg49\arcmin9.0\arcsec) \pm (0\arcdeg0\arcmin36.0\arcsec)_{stat} \pm (0\arcdeg1\arcmin12.0\arcsec)_{sys}$ \citep{Acciari2010}. The positions determined by \textit{Fermi}-LAT and VERITAS are consistent with each other, within statistical and systematic uncertainties, as well as with the center of the remnant.

In this work, we describe observations of Cas~A with two instruments: VERITAS and \textit{Fermi}-LAT. The main focus is on presenting the results of observations of Cas~A with VERITAS data taken between 2007 and 2013, which amount to more than 60 hours. This represents almost three times the previously published exposure by VERITAS, and significantly reduces the statistical errors on the flux, spectral index and centroid location. We perform extensive modelling using multiwavelength data available for Cas~A, and discuss different emission models for leptonic and hadronic scenarios.

\section{VERITAS: Observations and analysis results}
VERITAS (the Very Energetic Radiation Imaging Telescope Array System) is a ground-based $\gamma$-ray observatory which consists of an array of four telescopes, located in southern Arizona at an elevation of $1268 \U{m}$ above sea level \citep{Weekes2002, Holder2006}. Each telescope has a 12m-diameter optical reflector, providing a total reflecting area of $\sim 110 \U{m^{2}}$. The focal plane of each telescope is equipped with a camera consisting of $499$ photomultiplier tubes (PMTs) in a hexagonal close-packed array. The field of view of each PMT on the sky is $0.15^{\circ}$ in diameter, giving a total field of view of $3.5^{\circ}$ for each telescope. From 2007 to 2013, covering the period of data taking for Cas~A, the array underwent two major upgrades. The first occurred during the summer of 2009, when one telescope was relocated \citep{Perkins2009}. For the second upgrade, in summer 2012, all of the PMTs were replaced with new devices with a higher quantum efficiency \citep{Nepomuk2011, Kieda2013}. This improved the array sensitivity and lowered the energy threshold for observations. Currently, a source with a flux level of $1 \%$ of steady flux from the Crab Nebula can be detected in 25 hours. The angular resolution of the array at $1 \U{TeV}$ is $\sim$ $0.1^{\circ}$, and the sensitive energy detection range spans from $85 \U{GeV}$ to $30 \U{TeV}$ \citep{NaheeICRC2015}.

VERITAS observations of Cas~A are summarized in Table \ref{Observations}. Dataset I was taken between September $2007$ and November $2007$ with the original array configuration and, after data quality selection cuts, consists of $21$ hours of observations. Only $1.3$ hours of data (Dataset II) were taken between relocating one telescope and upgrading the camera, in December 2011. The total amount of good-quality data taken after the camera upgrade (Datasets III $\&$ IV) is $43$ hours. All data were taken in wobble mode \citep{Fomin1994}, in which a source is offset by $0.5^{\circ}$ (in this case) from the center of the field of view of the camera. This allows other regions, which do not contain the source, at the same radial distance from the camera center, to be used for estimating the background level. Data taken between September $2012$ and December $2013$ were divided in two parts; observations taken at small zenith angle (Dataset III) and large zenith angle (Dataset IV), with an average zenith angle of $31^{\circ}$ and $55^{\circ}$, respectively. Observations at large angles to the zenith result in a higher energy threshold, but with a larger effective collection area, boosting measurement of the highest-energy part of the source spectrum \citep{sommers1986}. In order to analyze this data, a standard VERITAS analysis procedure has been employed (for details see; \citet{Acciari2008, Cogan2008, MaierED2017}).

\begin{table*}
\caption{Details of VERITAS observations of Cas~A.}
\vspace{5mm}
\centering
\begin{tabular} {c c c c c c c}
\hline
\hline
Dataset& Date &Number of & Mean Zenith &Exposure Time & Previously\\
& & Telescopes & Angle (degree)&(Hours) & Published \\
\hline
I & 09/07 - 11/07 & 4 &34&21 & Yes \\
II & 12/11 - 12/11 & 4 &38&1.3 & No\\
III & 09/12 - 12/13 & 4 &31&20 & No\\
IV& 09/12 - 12/13 & 4 &55&23 & No\\
 
\hline
\end{tabular}
\label{Observations}
\end{table*}

The background was removed from the sample of $\gamma$-ray events using pre-determined cuts, which were optimized to give the best sensitivity for point-like sources with $3\%$ of the Crab Nebula flux. These cuts resulted in an energy threshold of $\sim 200 \U{GeV}$ for the Dataset presented here. Even after applying the cuts, there still existed some background, which was measured using the reflected region model \citep{Berge2007}. The significance of the source detection was calculated using Equation $17$ from \citet{LiMa1983}.

\subsection{Source localization and extension}
The best-fit centroid position of the emission from Cas~A in the energy range from 200 GeV to 8 TeV was measured by performing a maximum likelihood two-dimensional morphology fit using the \textit{Sherpa} package \citep{Sherpa2001}. For this analysis, two sky maps were used: (1) a count map of $\gamma$-ray like events containing both signal and background (\textit{ON} map), and, (2) a count map of $\gamma$-ray like background events (\textit{OFF} map) estimated using the reflected region model \citep{Berge2007}. In order to achieve the best angular resolution, only those events which were reconstructed using at least three telescope images, were selected. In addition, only small zenith angle data taken between 2012 and 2013 were used (Dataset III in Table \ref{Observations}).The statistical improvement achieved by adding large zenith angle and older data (Datasets I, II and IV) is offset by the additional systematic errors, which are significantly worse than for Dataset III ($\sim 70\arcsec$ in comparison to $\sim 25\arcsec$), and which exceed the statistical errors.


In the first step, the VERITAS PSF was determined using a reference source 1ES 1959+650. This is a blazar at a redshift of $z = 0.048$, which acts as a point-like source for VERITAS. Moreover, this source has similar declination and spectral shape compared to Cas A. Only data on 1ES 1959+650 taken under conditions similar to the Cas~A observations (same zenith angle, same array configuration) were selected. Under the assumption of point source, the signal events can be modelled by the VERITAS PSF, which is described by a two-dimensional King function, $k(r) = N_{0} (1 + (r/r_{0})^{2})^{-\beta}$, where $N_{0}$ is a normalization factor, $r$ is the angular distance from the centroid position, $r_{0}$ is the core radius, and $\beta$ is an index. By constraining the fitting range within $\pm 0.3^{\circ}$ region around the source of interest, the background events can be modelled with a two-dimensional constant function. In the first step of fitting, the background level was estimated by fitting a constant two-dimensional model to the \textit{OFF} map. In the second step, the constant 2D function plus a 2D King function was used to model the \textit{ON} map. During the fit in the second step, the parameters for the background model were frozen to the values calculated from the first step, while the centroid, $r_{0}$, and $\beta$ of the king function were allowed to vary. The best-fit source position of 1ES 1959+650 is measured as RA $= (19\hours59\minutes58.4\seconds) \pm (0\hours0\minutes1.8\seconds)_{stat} \pm (0\hours0\minutes3\seconds)_{sys}$ and Dec $= (+65\arcdeg9\arcmin35.6\arcsec) \pm (0\arcdeg0\arcmin10.7\arcsec)_{stat}\pm (0\arcdeg0\arcmin25\arcsec)_{sys}$, which is compatible with TeV catalogue position\footnote{\url{http://tevcat.uchicago.edu/?mode=1&showsrc=79}} at RA $= (19\hours59\minutes59.8\seconds)$ and Dec $= (+65\arcdeg8\arcmin55\arcsec)$. The $r_{0}$ and $\beta$ which define the PSF were calculated at a value of $0.094 \pm 0.014_{stat}$ degrees and $1.95 \pm 0.28_{stat}$ respectively.

\begin{figure}[htbp]
\begin{center}
\includegraphics[scale=0.5]{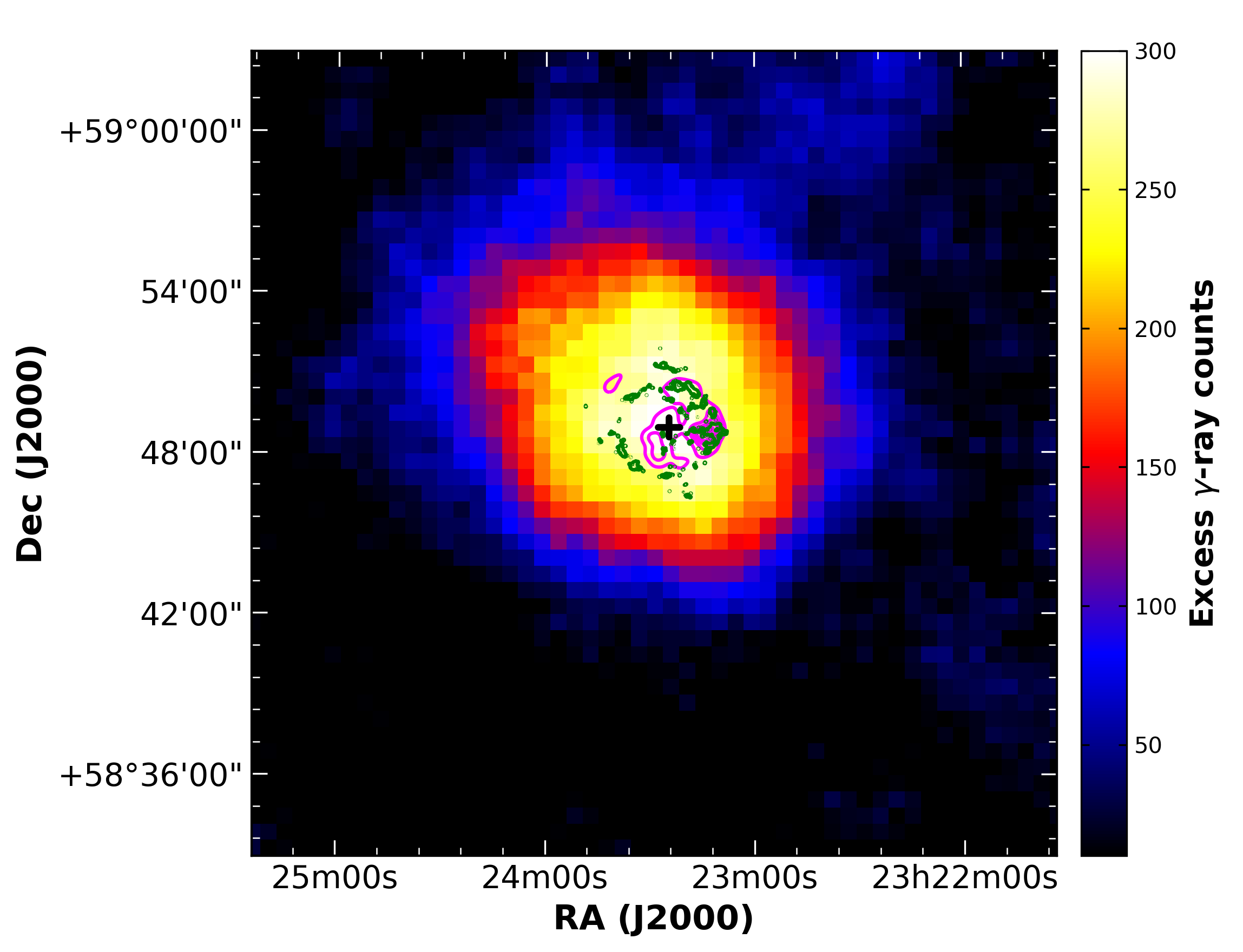}
\end{center}
\caption{The skymap of excess $\gamma$-ray events (with energy higher than 200 GeV) around the region of Cas~A, smoothed with a circular window of radius $0.09^{\circ}$. Magenta and green contours overlaid on this excess map are taken from \textit{NuStar} $15-20 \U{keV}$ X-ray emission \citep{Grefenstette2014} and the VLA $6\U{cm}$ radio emission respectively. The black cross indicates the measured centroid position of the $\!\!\U{TeV}$ $\gamma$-ray source.}
\label{fig:CasAPositionFitting}
\end{figure}

Similar analysis procedure was followed as above to get the source position for Cas~A, under the assumption that it is an unresolved source for the VERITAS. However, the $\beta$ parameter was fixed to the a value calculated from the analysis on 1ES 1959+650. We find the best-fit source position, in equatorial coordinates at RA $= (23\hours23\minutes24.4\seconds) \pm (0\hours0\minutes3.2\seconds)_{stat} \pm (0\hours0\minutes3\seconds)_{sys}$ and Dec $= (+58\arcdeg48\arcmin59.1\arcsec) \pm (0\arcdeg0\arcmin23.0\arcsec)_{stat} \pm (0\arcdeg0\arcmin25\arcsec)_{sys}$. This best-fit source position for Cas A is shown as black cross on Figure \ref{fig:CasAPositionFitting} which shows the skymap of excess $\gamma$-ray counts from the region of Cas~A, smoothed with a circular window of radius $0.09^{\circ}$. This map was produced using 20 hours of VERITAS observations from $2012$ (with the upgraded camera and at small zenith angles). Based on fitting results, the TeV $\gamma$-ray source in the region of Cas~A is named VER J2323+588. The $r_{0}$ is found to be $0.084\pm0.008_{stat}$ degrees. This is compatible with the $r_{0}$ value for reference source 1ES 1959+650 within $2\sigma$ statistical errors. This indicates that the position of centroid and the point-like source nature of VER J2323+588 are consistent with the origin of the emission being from the Cas~A SNR. The magenta and green contours taken from \textit{NuStar} $15-20 \U{keV}$ X-ray emission \citep{Grefenstette2014} and the VLA $6\U{cm}$ (Courtesy of DeLaney\footnote{\url{http://homepages.spa.umn.edu/~tdelaney/cas/}}) radio image respectively are also overlaid on this excess map, which shows that the centroid of gamma-ray emission lies within the radio and X-ray extent of SNR Cas~A.



\subsection{Spectral analysis}

To derive the energy spectrum, the entire Dataset (I, II, III and IV) was used. A total of $1535$ $\gamma$-ray like events ($N_{on}$) were counted from a region of radius $0.09^{\circ}$ around Cas~A. Since this region also contain background events, background is obtained by counting the total number of events from $6$ identical source-free regions in the same field of view using reflected region model \citep{Berge2007}. This gives $N_{off} = 6241$. By taking into account the ratio of area of \textit{on} and \textit{off} regions ($\alpha=0.167$), excess number of $\gamma$-ray events were calculated at a value of  $N_{excess} = 495 \pm 41$. The significance of this detection, calculated using Equation 17 of \citet{LiMa1983}, was $13.1 \sigma$. The excess $\gamma$-ray events are then binned into $9$ equal  logarithmically spaced energy bins to obtain the differential energy spectrum (see Table \ref{SedVeritas}). Above the threshold energy of $200 \U{GeV}$, the spectrum is well-described by a PL distribution (See Figure \ref{fig:BroadBandSED}):
\begin{equation}
\frac{dN} {dE} = (1.45 \pm 0.11_{stat}) \times 10^{-12} ({E} / {\rm{TeV}})^{-2.75 \pm 0.10_{stat}} \rm{~cm^{-2}~s^{-1}~TeV^{-1}}\,.
\end{equation}

A PL fit to the data points gives a $\chi^{2}$ of 2.2 for 5 degrees of freedom, resulting in a fit probability of 81\%. This result is in agreement with the previously published HEGRA \citep{Aharonian2001}, VERITAS \citep{Acciari2010} and  MAGIC \citep{Albert2007} spectral measurements, when both statistical and systematic errors are taken into account. Compared to previously published VERITAS spectral results \citep{Acciari2010}, the present work leads to a reduction of statistical errors on the spectral index and flux normalization by $ \sim 60 \%$ and $\sim 40\%$ respectively. The differential flux points measured by VERITAS (see Table \ref{SedVeritas}) are also compatible with the recent MAGIC results \citep{Ahnen2017}. 

\begin{table}[htbp]
\caption{Differential spectral flux points with statistical errors from VERITAS data in the energy range $0.2-12.6 \U{TeV}$. Upper limits of differential flux are obtained at $95\%$ confidence level for those points where significance is less than $2\sigma$.}
\vspace{5mm}
\centering
\begin{tabular}{ ccccc  }
\hline
\hline
Energy & Energy min & Energy max & $\mathrm {E^{2} dN(E)/dE}$ & Significance \\
($\mathrm{TeV})$ & ($\mathrm{TeV}$) & ($\mathrm{TeV}$) & ($\mathrm{10^{-12}\,erg^{1}\,s^{-1}\,cm^{-2}}$) & ($\sigma$)\\
\hline
0.25 & 0.20 & 0.32	&15.20 (upper limit) & 0.1 \\[1ex]
0.40 & 0.32 & 0.50	& $4.28_{-0.76}^{+0.79}$ & 6.1 \\[1ex]
0.63 & 0.50 & 0.79	& $3.64_{-0.51}^{+0.53}$ & 8.1 \\[1ex]
1.00 & 0.79 & 1.26	& $2.17_{-0.39}^{+0.40}$ & 6.3 \\[1ex]
1.58 & 1.26 & 2.00 & $1.83_{-0.35}^{+0.37}$ & 5.9 \\[1ex]
2.51 & 2.00 & 3.16 & $1.40_{-0.31}^{+0.34}$ & 5.3 \\[1ex]
3.98 & 3.16 & 5.01 & $0.63_{-0.26}^{+0.29}$ & 2.7 \\[1ex]
6.31 & 5.01 & 7.94 & $0.50_{-0.21}^{+0.25}$ & 2.8 \\[1ex]
10.00 & 7.94 & 12.59 & 0.37 (upper limit) & 1.0  \\[1ex]
\hline

\end{tabular}
\label{SedVeritas}
\end{table}

\section{\textit{Fermi}-LAT: Observations and analysis results} 
The LAT instrument on board the \textit{Fermi} satellite is a pair-conversion $\gamma$-ray detector that detects photons in the energy range between 20\U{MeV} and $>$ 500\U{GeV}. The LAT has a field of view of $\sim$2.4\U{sr}, effective area of $\sim 8200 \U{cm^{2}}$ on-axis above $1 \U{GeV}$ (\textit{Pass 8} events) and an angular resolution of $\sim 0.8^{\circ}$ at $1 \U{GeV}$. Full details about the LAT instrument can be found in \cite{Atwood2009}. 

We analyzed 10.8 years of \textit{Pass 8 R3} LAT data (see \citet{Atwood2013} for more details), from 2008 August 4 to 2019 May 31. We used the \textit{Fermipy}\footnote{\url{http://fermipy.readthedocs.org/en/latest/}} Python package (version 0.17.4, \citet{WoodFermipy2017}) that automates the analysis of \textit{Pass 8} data in conjuction with  the publicly available software \textit{fermitools}, version 1.0.1. We selected events from a $20^{\circ} \times 20^{\circ}$ region centered on the position of Cas~A in the energy range from $100\U{MeV}$ to $500\U{GeV}$. In order to minimize the contamination from cosmic rays mis-classified as $\gamma$-rays, we selected events belonging to the \textit{UltraCleanVeto} Class ($evclass=1024$). Data were filtered further by selecting only \textit{PSF2} and \textit{PSF3} ($evtype=16$ and $32$) event types that give the best angular resolution. For details about the event classes and event types see the \textit{Fermi} web pages\footnote{\url{https://fermi.gsfc.nasa.gov/ssc/data/analysis/documentation/Cicerone/Cicerone\_Data/LAT\_DP.html}}. Once this data selection was made, we applied another cut to select the good time intervals by using (\texttt{DATA\_QUAL}) $> 0$ \&\& \texttt{(LAT\_CONFIG} $==1$). In order to avoid the contamination from photons produced by cosmic ray interactions in the upper atmosphere, we applied a zenith angle cut of $ \theta < 90^{\circ}$. The remaining photons were binned using the \textit{gtbin} tool into a spatial bin size of $0.1^{\circ} \times 0.1^{\circ}$ and into 22 equal logarithmically-spaced energy bins. 

We applied the likelihood technique to find the parameters of the source of interest, where likelihood is defined as the probability of data given the model. A joint likelihood function was defined in this work by taking the product of the likelihood function of \textit{PSF2} and \textit{PSF3} type events. The maximization of this likelihood function provided the parameters of the input model. The input model file used in the binned likelihood analysis was created by including all of the background sources within $20^{\circ}$ from the center of the region of interest (ROI) from the 4FGL catalogue \citep{4FGL2019}. In addition to this, two background diffuse models; Galactic (\textit{gll\_iem\_v07.fits}) and extragalactic (\textit{iso\_P8R3\_ULTRACLEANVETO\_V2\_PSF2\_v1.txt}, \textit{iso\_P8R3\_ULTRACLEANVETO\_V2\_PSF3\_v1.txt}) were also included in the input model, and the normalization was set free for these two models. During the maximum-likelihood fitting of data with \textit{gtlike}, the normalization and spectral parameters of sources within $3^{\circ}$ from the center of the ROI were set free. The parameters for other sources, located outside of the $3^{\circ}$ radius, were fixed and set at their catalogue values. The instrument response function (IRF) used in our analysis was \textit{P8R3\_ULTRACLEANVETO\_V2}.

\subsection{Source localization and extension}
For source localization in the high-energy band, we selected \textit{P8R3 SOURCE} class with front plus back type $\gamma$-ray events in the energy range from $10 \U{GeV} \leq E_{\gamma} \leq 500 \U{GeV}$. Such a selection provides a good instrument PSF ($\sim 0.1^{\circ}$) and less contamination from the Galactic diffuse emission which dominates below $1 \U{GeV}$. The best-fit source position in Galactic coordinates was obtained by the \textit{Source Localization} routine in the \textit{Fermipy} package. This routine uses a two step method to find the best-fit source position. In the first step, a likelihood map of size $1^{\circ} \times 1^{\circ}$ around the known position of Cas~A is generated, and a fit is performed to find the position of the peak likelihood in the map. This position is further refined in the second step by freeing the location parameters of Cas~A and redoing the likelihood fitting in a smaller region that encloses the $99\%$ positional uncertainty contour from the first step. The result of this localization analysis gave the best-fit position at RA $= 23\hours23\minutes26.5\seconds$ and Dec $= +58\arcdeg48\arcmin59.8\arcsec$, with a $1 \sigma$ statistical uncertainty of $0.2 \arcmin$. This new position is offset from the previous position given in \cite{Yuan2013} by $0.5 \arcmin$, but is compatible with this result because the systematic error in the position due to the alignment of the telescope system and inaccurate description of the PSF of the instrument is estimated to be $0.3 \arcmin$. We also performed an extension analysis of the source using the \textit{source extension} routine in the \textit{Fermipy} package. We tested the extension of the source by comparing the likelihood of the extended source hypothesis to the point source hypothesis. For the extended source hypothesis, we tested two source morphology models; a 2D symmetric Gaussian model and a radial disk model. Both models yield no significant detection of the extension. With a confidence level of $95\%$, we calculate the upper limit of the source extension to be $2.2 \arcmin$ and $2.5 \arcmin$ with the 2D Gaussian model and the radial disk model, respectively. These values for the upper limit on the extension are consistent with the size of the SNR ($2.55 \arcmin \pm 0.2  \arcmin$ \citet{Gotthelf2001}).

\subsection{Spectral analysis}
The spectral analysis was performed over the full \textit{Fermi}-LAT energy range of $0.1-500 \U{GeV}$ using \textit{gtlike}. Following \citet{Yuan2013}, the spectral shape of the emission from Cas~A was assumed to be smoothly broken power-law (SBPL; $dN/dE = N_{0} (E/E_{0})^{-\gamma_{1}} (1 + (E/E_{b})^{\frac{\gamma_{2} - \gamma_{1}}{\beta}})^{-\beta}$, where $N_{0}$ is the normalization factor, $E_{0}$ is the scale parameter fixed at a value of $1 \U{GeV}$, $E_{b}$ represents the break energy in the spectrum, $\gamma_{1}$ and $\gamma_{2}$ are the photon indexes before and after the break, and $\beta$ represents the smoothness of the break and is fixed to $0.1$). The parameters for the SBPL model are shown in Table \ref{SBPLModelParameters} and the differential flux points are shown in Table \ref{SedFermi}.






\begin{table}[htbp]
\caption{\textit{Fermi-LAT} results: smoothly broken power-law model parameters with statistical error.}
\vspace{5mm}
\centering
\begin{tabular}{ cccccc  }
\hline
\hline
$N_{0}$ & $E_{0}$ & $\gamma_{1}$ & $\gamma_{2}$ & $E_{b}$ & $\beta$ \\
($\rm cm^{-2}~s^{-1}~MeV^{-1}$) & (GeV) & & & (GeV) & \\
\hline
$(6.4 \pm 0.7) \times 10^{-12}$ & 1.0 (fixed) & $1.3\pm 0.2$ & $2.1\pm 0.1$ & $1.3 \pm 0.4$ & 0.1 (fixed) \\[1ex]

\hline

\end{tabular}
\label{SBPLModelParameters}
\end{table}

For calculating the SED, the energy range from $0.1$ to $500 \U{GeV}$ was divided into $22$ logarithmically spaced bins. We used the \textit{sed} method in the \textit{Fermipy} package, where SED is computed by performing fitting of flux of Cas~A  in each energy bin independently, using a fixed spectral index of 2, intermediate between the two indices obtained for the smoothly broken power-law fit and consistent with the index obtained from the global fit to a simple power-law. We determined that the resulting SED flux points (given in Table \ref{SedFermi}) are insensitive to this choice of index by fitting with indices 1.3 and 2.1 instead and finding the points to differ by less than their error bars. 

\begin{figure}[htbp]
\begin{center}
\includegraphics [width=14cm,height=10cm]{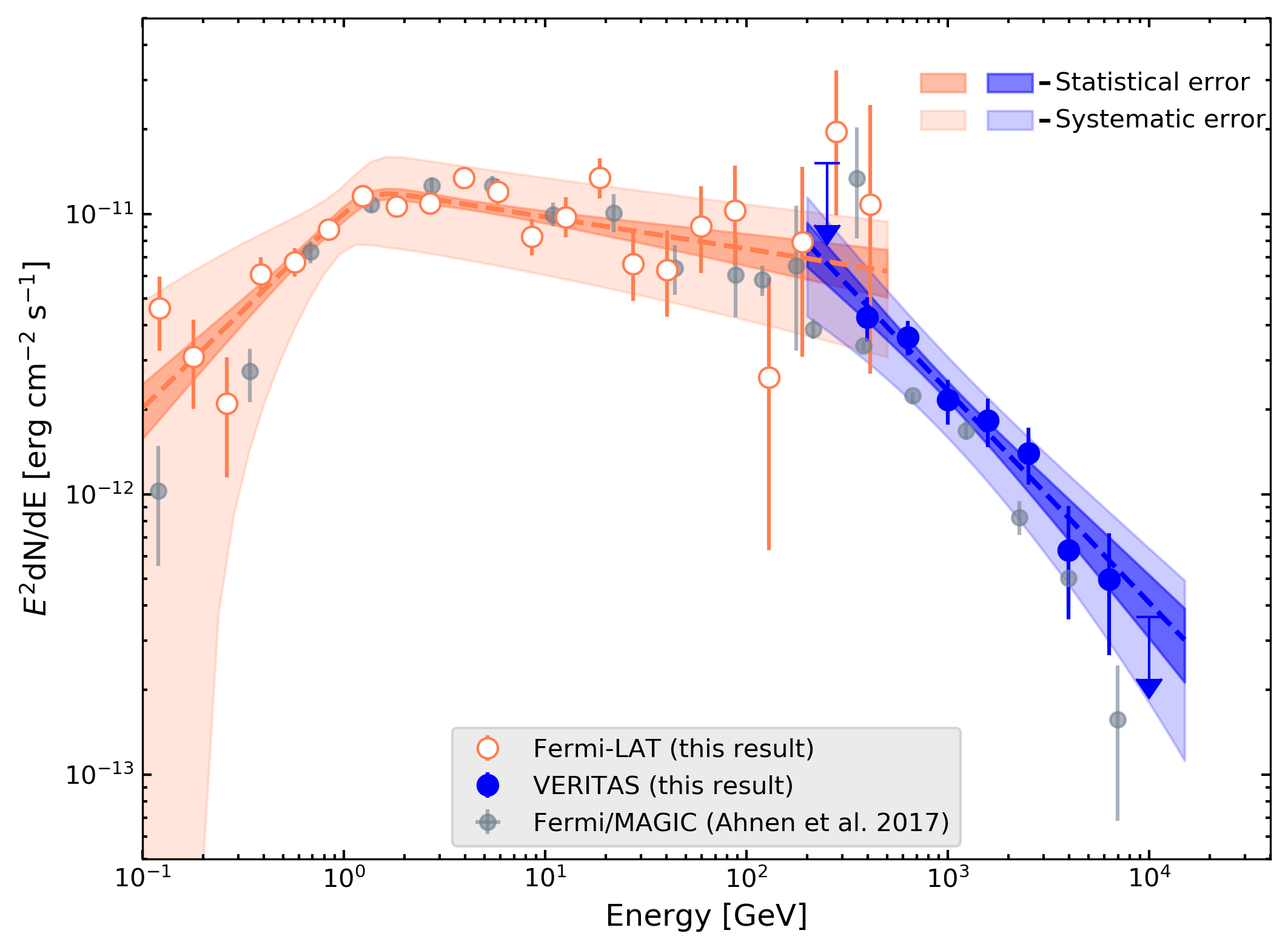}
\end{center}
\caption{Broadband SED of Cas~A using \textit{Fermi}-LAT and VERITAS points. For comparison, \textit{Fermi}-LAT/MAGIC SED points measured in \citet{Ahnen2017} are also plotted in grey. Orange (blue) shaded region represents the $1 \sigma$ statistical error band on the spectral fit of \textit{Fermi}-LAT (VERITAS). Similarly, the light-orange (light-blue) shaded region represents $1 \sigma$ systematic errors (only; not including statistical errors) for \textit{Fermi}-LAT (VERITAS). \textit{Fermi}-LAT points (open orange circles) are fitted with SBPL from $0.1-500 \U{GeV}$ and VERITAS points (filled blue circles) are fitted with simple power-law from $200-15000 \U{GeV}$. }
\label{fig:BroadBandSED}
\end{figure}

In the fitting process, the normalization of the Galactic diffuse model was also allowed to vary. Table \ref{SedFermi} shows the differential flux points in all bins. As mentioned in \citet{Yuan2013}, the uncertainty of the modeling of Galactic diffuse emission is the major contribution for the systematic error on the spectral measurements. Therefore, we consider the impact of this component to overall spectrum measurement. To estimate this error, we calculated the discrepancy between the number of counts  predicted from the best-fit model and the data at 17 random locations close to the position of Cas~A, but away from all known sources (similar to the procedure adopted in \cite{Abdo2009a}). The differences between the best-fit model and data were found to be $\sim 5\%$. In order to estimate the systematic error, therefore, we changed the normalization of the Galactic diffuse model artificially by $\pm 5\%$ from the best-fit values. Figure \ref{fig:BroadBandSED} shows the Cas~A SED from \textit{Fermi}-LAT data with systematic and statistical errors. For comparison, SED points from \citet{Ahnen2017}, measured by the MAGIC collaboration, are also plotted on  Figure \ref{fig:BroadBandSED}. All of the SED points from this work are consistent with the published \citet{Ahnen2017} points within 1-2 $\sigma$ considering both statistical and systematic errors . 

\begin{table}[htbp]
\caption{SED points from \textit{Fermi}-LAT data in the energy range $0.1-500 \U{GeV}$ (only statistical errors).}
\vspace{5mm}
\centering
\begin{tabular}{ ccccc  }
\hline
\hline
Energy & Energy min & Energy max & $\mathrm {E^{2} dN(E)/dE}$ & Significance \\
($\mathrm{GeV})$ & ($\mathrm{GeV}$) & ($\mathrm{GeV}$) & ($\mathrm{10^{-12}\,erg^{1}\,s^{-1}\,cm^{-2}}$) & ($\sigma$)\\
\hline
0.12 & 0.10 & 0.15 & $4.60_{-1.35}^{+1.36}$ & 3.4 \\[1ex]
0.18 & 0.15 & 0.22 & $3.10_{-1.08}^{+1.09}$ & 2.9 \\[1ex]
0.26 & 0.22 & 0.32 & $2.11_{-0.96}^{+0.97}$ & 2.2 \\[1ex]
0.39 & 0.32 & 0.47 & $6.09_{-0.88}^{+0.90}$ & 7.2 \\[1ex]
0.57 & 0.47 & 0.69 & $6.74_{-0.78}^{+0.80}$ & 9.4 \\[1ex]
0.84 & 0.69 & 1.02 & $8.81_{-0.75}^{+0.77}$ & 13.9 \\[1ex]
1.24 & 1.02 & 1.50 & $11.60_{-0.79}^{+0.81}$ & 19.7 \\[1ex]
1.82 & 1.50 & 2.21 & $10.60_{-0.78}^{+0.82}$ & 19.4 \\[1ex]
2.69 & 2.21 & 3.26 & $10.90_{-0.85}^{+0.90}$ & 20.8 \\[1ex]
3.96 & 3.26 & 4.80 & $13.40_{-1.05}^{+1.13}$ & 24.0 \\[1ex]
5.83 & 4.80 & 7.07 & $12.00_{-1.21}^{+1.31}$ & 19.4 \\[1ex]
8.58 & 7.07 & 10.41 & $8.31_{-1.18}^{+1.30}$ & 14.6 \\[1ex]
12.64 & 10.41 & 15.34 & $9.75_{-1.50}^{+1.71}$ & 15.0 \\[1ex]
18.61 & 15.34 & 22.59 & $13.40_{-2.08}^{+2.37}$ & 16.3 \\[1ex]
27.41 & 22.59 & 33.27 & $6.62_{-1.73}^{+2.05}$ & 9.2 \\[1ex]
40.37 & 33.27 & 49.00 & $6.30_{-2.01}^{+2.43}$ & 7.7 \\[1ex]
59.46 & 49.00 & 72.16 & $9.04_{-2.88}^{+3.49}$ & 8.3 \\[1ex]
87.57 & 72.16 & 106.27 & $10.30_{-3.74}^{+4.62}$ & 7.4 \\[1ex]
128.97 & 106.27 & 156.52 & $2.61_{-1.98}^{+3.34}$ & 2.6 \\[1ex]
189.95 & 156.52 & 230.52 & $7.93_{-4.84}^{+6.78}$ & 3.6 \\[1ex]
279.75 & 230.52 & 339.50 & $19.60_{-9.74}^{+12.80}$ & 6.0 \\[1ex]
412.01 & 339.50 & 500.00 & $10.80_{-8.11}^{+13.60}$ & 3.3 \\[1ex]
\hline

\end{tabular}
\label{SedFermi}
\end{table}

\section{Combined \textit{Fermi}-LAT and VERITAS results}

\subsection{Centroid positions}
\begin{figure}[htbp]
\begin{center}
\includegraphics [width=10cm,height=8.5cm]{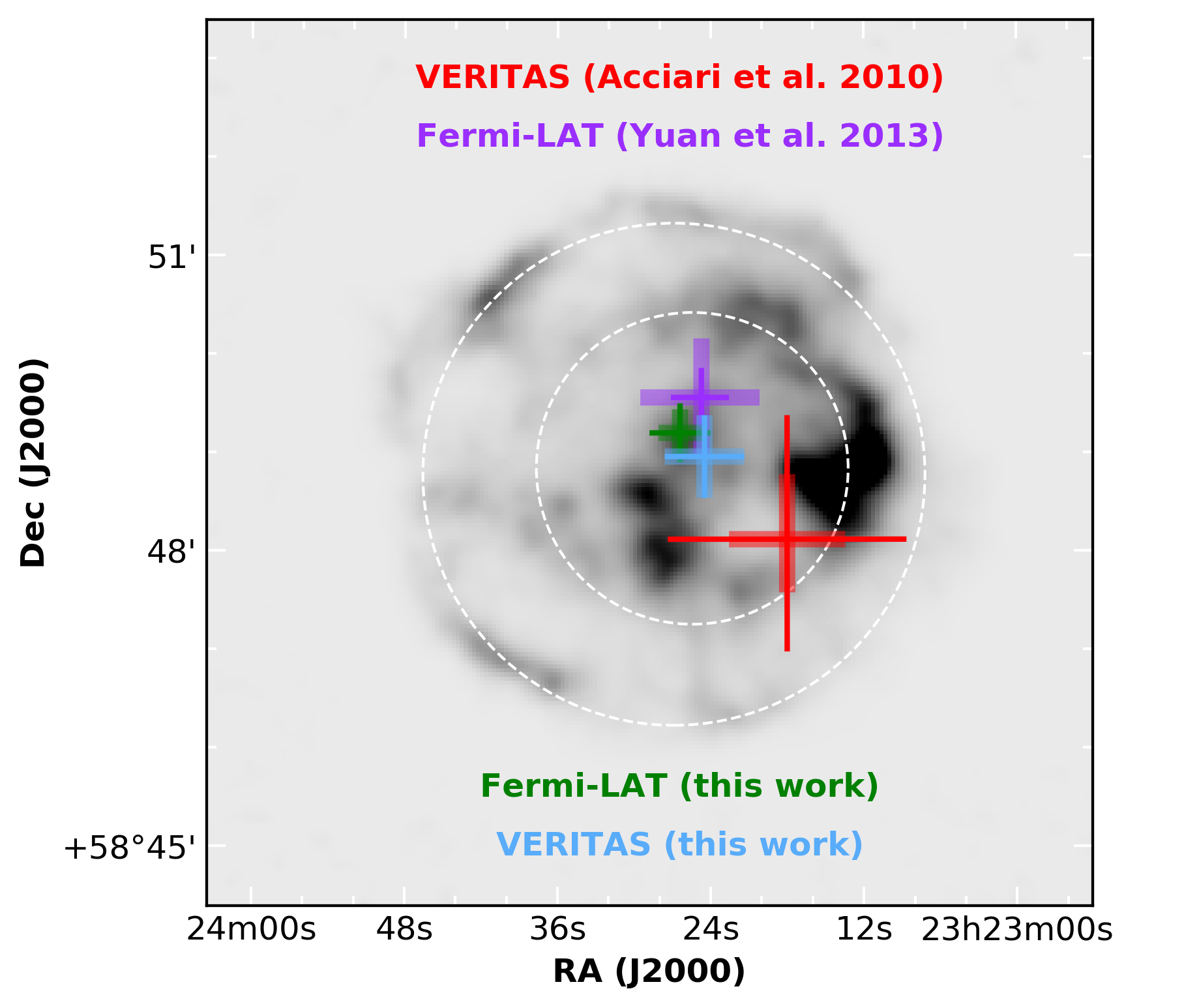}
\end{center}
\caption{Comparison of $\U{GeV}$ and $\U{TeV}$ centroid positions. The background image shows the \textit{NuSTAR} $15-20 \U{keV}$ hard X-ray emission from Cas~A \citep{Grefenstette2014}. The two dashed circles denote the positions of forward and reverse shocks \citep{Gotthelf2001}. Updated VERITAS (for $\gamma$-ray above $200 \U{GeV}$ energy) and \textit{Fermi}-LAT (for $\gamma$-ray above $10 \U{GeV}$ energy) centroid positions are denoted by green and blue crosses. The thick crosses represent $1 \sigma$ statistical errors and thin crosses represent $1 \sigma$ systematic errors. Also shown here are the best-fit positions from the previous VERITAS \citep{Acciari2010} and \textit{Fermi}-LAT observations \citep{Yuan2013} in red and purple crosses, respectively.}
\label{fig:CentroidPositions}
\end{figure}

Figure \ref{fig:CentroidPositions} shows the hard X-ray emission from Cas~A measured using \textit{NuSTAR} telescope in the energy range $15-20 \U{keV}$ \citep{Grefenstette2014} with the centroid positions of the $\!\!\U{GeV}$ and the $\!\!\U{TeV}$ emission. The best-fit positions obtained with the \textit{Fermi}-LAT and VERITAS are compatible with each other and lie close to the center of the remnant. Because the PSF of \textit{Fermi}-LAT and VERITAS is comparable to the size of remnant, it is difficult to compare the emission locations for hard X-rays, GeV and TeV $\gamma$-rays. Therefore, a morphological comparison between hard X-ray emission and $\gamma$-ray emission does not help us to interpret the emission mechanism for GeV and TeV $\gamma$-rays at this point. 

\subsection{Broadband spectral fit}

\begin{figure}[htbp]
\begin{center}
\includegraphics [width=12cm,height=10cm]{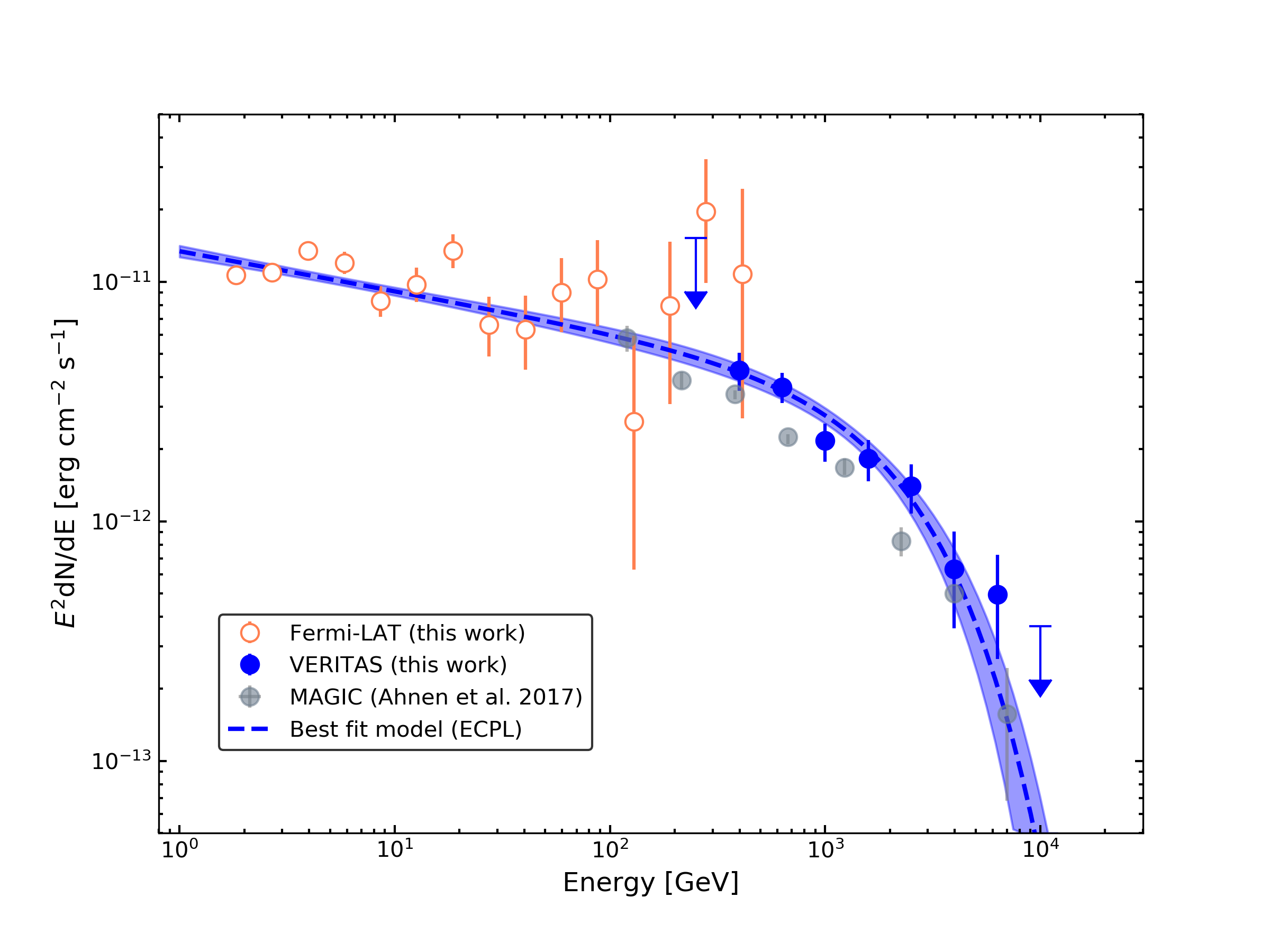}
\end{center}
\caption{\textit{Fermi}-LAT and VERITAS measured spectral points of Cas~A. Assuming only statistical errors, the best-fit ECPL model is shown with a dotted blue line. The blue shaded region represents the $1 \sigma$ statistical error band on the best-fit ECPL spectral model.}
\label{fig:FermiVeritas}
\end{figure}

Up to this point, we have calculated the flux points from \textit{Fermi}-LAT and VERITAS data independently using different analysis packages. Here, we take those flux points, assuming that they are independent, and combine them to perform a broadband spectral fit. We performed the broadband fit above the break energy of the \textit{Fermi}-LAT spectrum to check the spectral behaviour at the highest end of the energy range. The spectral points from the \textit{Fermi}-LAT (above the spectral break only, i.e. $> 1.3 \U{GeV}$) and VERITAS are fitted jointly using three different models: a single PL, an exponential cut-off power-law (ECPL) and a SBPL. See Table \ref{DifferentPowerLaws} for the formula of each spectral model. The PL fit yields a $\chi^{2}$-fit probability of $3.1 \times 10^{-7}$, whereas the ECPL and SBPL yield  $\chi^{2}$-fit probabilities of 0.06 and 0.13, respectively. The ECPL and SBPL models are therefore favored over the PL model at the $6.0\sigma$ level when only statistical errors are considered. Adding a systematic error of $0.1_{sys}$ \citep{Yuan2013} in the \textit{Fermi} spectral index and $0.2_{sys}$ \citep{MadhavanThesis2013} on the VERITAS spectral index, reduces the significance of the ECPL and SBPL over PL to $\sim 4.0\sigma$ level. Since both ECPL and SBPL show similar significance, and ECPL has fewer parameters than SBPL, we take ECPL as the best-fit model for our Dataset. Figure \ref{fig:FermiVeritas} shows the best-fit ECPL model on the joint \textit{Fermi}-LAT and VERITAS spectral points. The energy of the cut-off is measured to be $2.3 \pm 0.5_{stat} \U{TeV}$. This value is consistent with the cutoff of $3.5_{-1.0}^{+1.6} \U{TeV}$ measured by MAGIC \citep{Ahnen2017}.

\begin{table}[htbp]
\caption{Comparison of different spectral models for the fit to the \textit{Fermi}-LAT and VERITAS data above $1.3 \U{GeV}$.}
\vspace{5mm}
\centering
\begin{tabular}{ llll  }
\hline
\hline
Spectral Model & Formula &Parameter values & $\chi^{2}$ / ndf \\
\hline \\

\multirow{1}{*}{PL} & $N_{0}(E/E_{0})^{-\gamma}$ & $\gamma = 2.30 \pm 0.01$  & 68/20 \\\\
\hline \\
\multirow{2}{*}{ECPL} & \multirow{2}{*}{$N_{0}(E/E_{0})^{-\gamma} \exp(-E/E_{c})$} & $\gamma = 2.17 \pm 0.02$  & \multirow{2}{*}{30/19} \\ 
& & $E_{c}\:(\!\!\U{TeV}) = 2.31 \pm 0.51$ &  \\ \\
\hline\\

\multirow{3}{*}{SBPL} &\multirow{3}{*}{$N_{0} (E/E_{0})^{-\gamma_{1}} (1 + (E/E_{b})^{\frac{\gamma_{2} - \gamma_{1}}{\beta}})^{-\beta}$} & $\gamma_{1} = 2.11 \pm 0.04$  & \multirow{3}{*}{25/18} \\ 
 & & $\gamma_{2} = 2.77 \pm 0.10$ &  \\
 & & $E_{b}\:(\!\!\U{TeV}) = 0.25 \pm 0.09$ &  \\\\
\hline

\end{tabular}
\label{DifferentPowerLaws}
\end{table}

\section{Theoretical modeling}
\subsection{Model assumptions}
We build a global model to investigate the multi-wavelength spectrum from radio up to the TeV energy range. For simplicity, we assume a one-zone model fixed by two parameters: the ambient hydrogen number density, $n_{\mathrm{H}}$, and the post-shock magnetic-field strength, $B$. Both quantities are assumed constant, i.e. independent of time and location. The differential electron (proton) number densities, $N_{e(p)}$, are assumed to follow ECPL
\begin{equation}
 N(p)=N_{0}\, p^{-s}\exp\left(-\frac{p}{p_{cut}}\right)\,.
 \label{eq:spectra}
\end{equation}
Here $p$, $p_{cut}$, and $s$ denote the electron (proton) momentum, the cut-off momentum and the PL index of the spectrum, respectively, all of which are free parameters of our model. The normalization, $N_0$, in principle reflects the injection efficiency of each particle species. We calculate the synchrotron emission from the non-thermal electron spectrum \citep{synchrotron&IC}, taking into account the modifications caused by the turbulent component of the magnetic field \citep{reacc}. NTB and IC radiation, which can significantly contribute to the $\gamma$-ray spectrum of SNR, are also obtained from the non-thermal electron distribution. For the IC interactions \citep{synchrotron&IC}, we consider two target photon fields: the cosmic microwave background and the infrared emission from the shock-heated ejecta with temperature $\sim 100\, \mathrm{K}$ and energy density $2\,\mathrm{eV\, cm}^{-3}$ \citep{Mezger}. The NTB contribution from relativistic electrons follows the calculations of \citet{synchrotron&IC}. Additionally, thermal bremsstrahlung from plasma electrons is included assuming local thermodynamic equilibrium \citep{thermal}. The $\gamma$-ray yield from protons via neutral-pion decay is computed using the procedure of \citet{gammaProt}. Including the hydrogen number density and the magnetic field strength, we have in total nine independent parameters in our global model. The parameters are shown in Table~\ref{Theor_param}. The hydrogen number density, $n_{\mathrm{H}}$, corresponds to the upstream value and magnetic-field strength, $B$, to the downstream region. In the following, we consider two scenarios: a hadron-dominated model and a lepto-hadronic case, which we refer to as Model~I and II, respectively.  

\begin{table}[!phtb]
 \caption{Parameters for theoretical models.}
 \vspace{5mm}
 \centering
 \begin{tabular}{c c c c c c | c c c c}
 \hline
 \hline
 &\multicolumn{5}{c|}{ Varying parameters} &\multicolumn{4}{c}{Same for both models}\\
 \hline
 Model & $B$  & $N_{0,e}$& $N_{0,p}$ &$p_{cut,e}$   &$p_{cut,p}$ &$s_{e}$ &$s_{p}$& $T_e$ & $n_{\mathrm{H}}$\\
 &  ($\mathrm{ \mu G}$) &$(\mathrm{(m_{e}c)^{s_e-1}})$ & $(\mathrm{(m_{p}c)^{s_p-1}})$&($\mathrm{m_{e}c}$) &($\mathrm{m_{p}c}$) & & &($10^{7}$K)& ($\mathrm{cm^{-3}}$)\\
 \hline
 I  & 450  & $4.2\times10^{13}$ & $3.2\times 10^{23}$& $9.0\times10^6$  & $2.1\times 10^{4}$ & 2.5 & 2.17 & 1.8 & 1.0\\
 II  & 150 & $2.9\times 10^{14}$ &$3.8\times 10^{23}$& $1.6\times10^{7}$  &$6.0\times 10^{3}$& 2.5 & 2.17 & 1.8 & 1.0\\
 \hline
 \end{tabular}
 \label{Theor_param}
 \end{table}

\subsection{Hadronic model}\label{sec:had_mod}
\begin{figure}[!phtb]
\begin{center}
\includegraphics[scale=0.6]{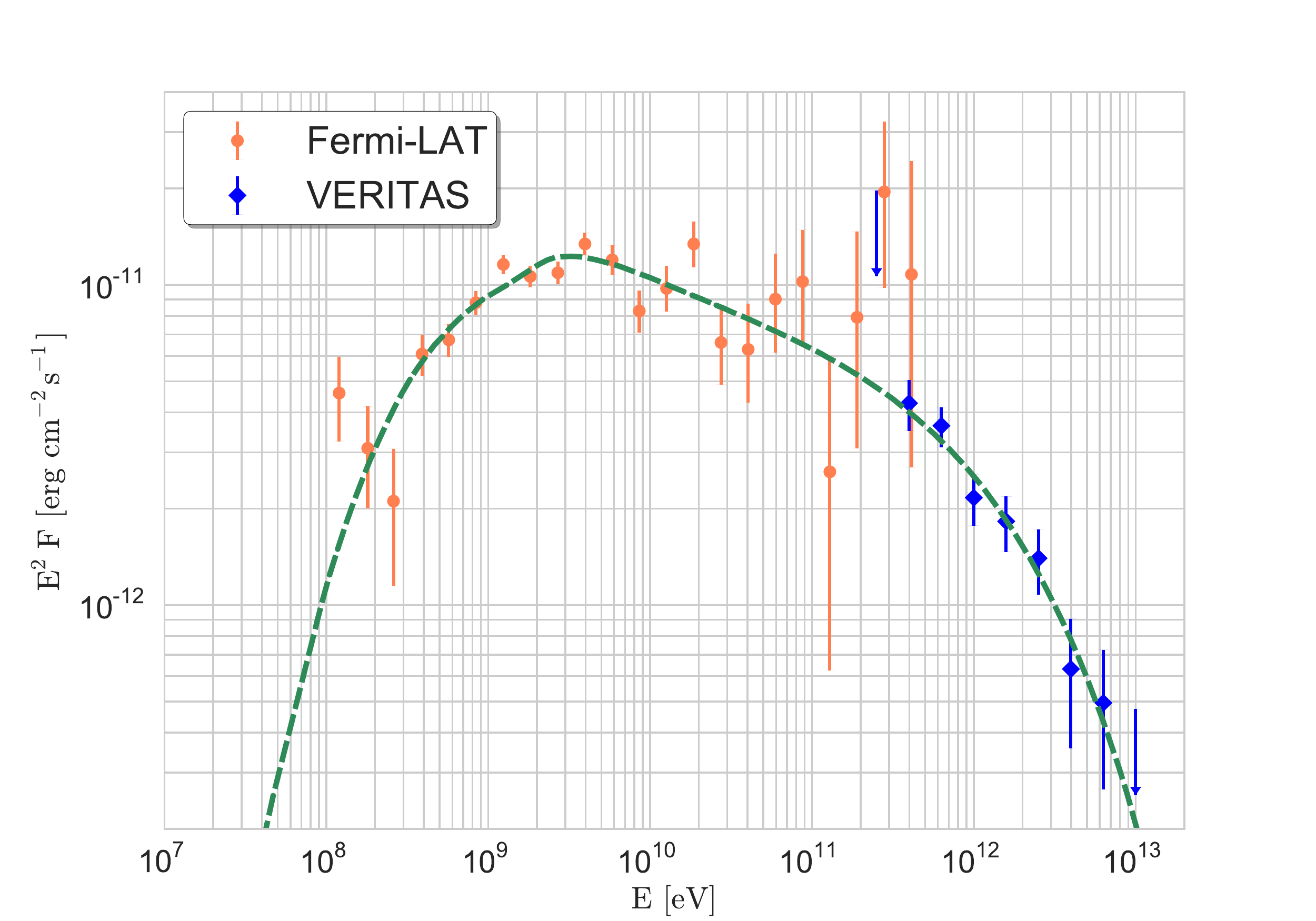}
\caption{\protect \small Purely hadronic best-fit with $\chi^2=36.01$ and $d.o.f.=26$ ($\chi^2/d.o.f.=1.38$). The corresponding best-fit parameters following Equation.~\ref{eq:spectra} are $s_p=2.17$ and $p_{cut}=2.1\times 10^4\,\mathrm{m_{p}c}$ .}
\label{hadbestfit}
\end{center}
\end{figure}

\begin{figure}[!phtb]
\begin{center}
\includegraphics[scale=0.6]{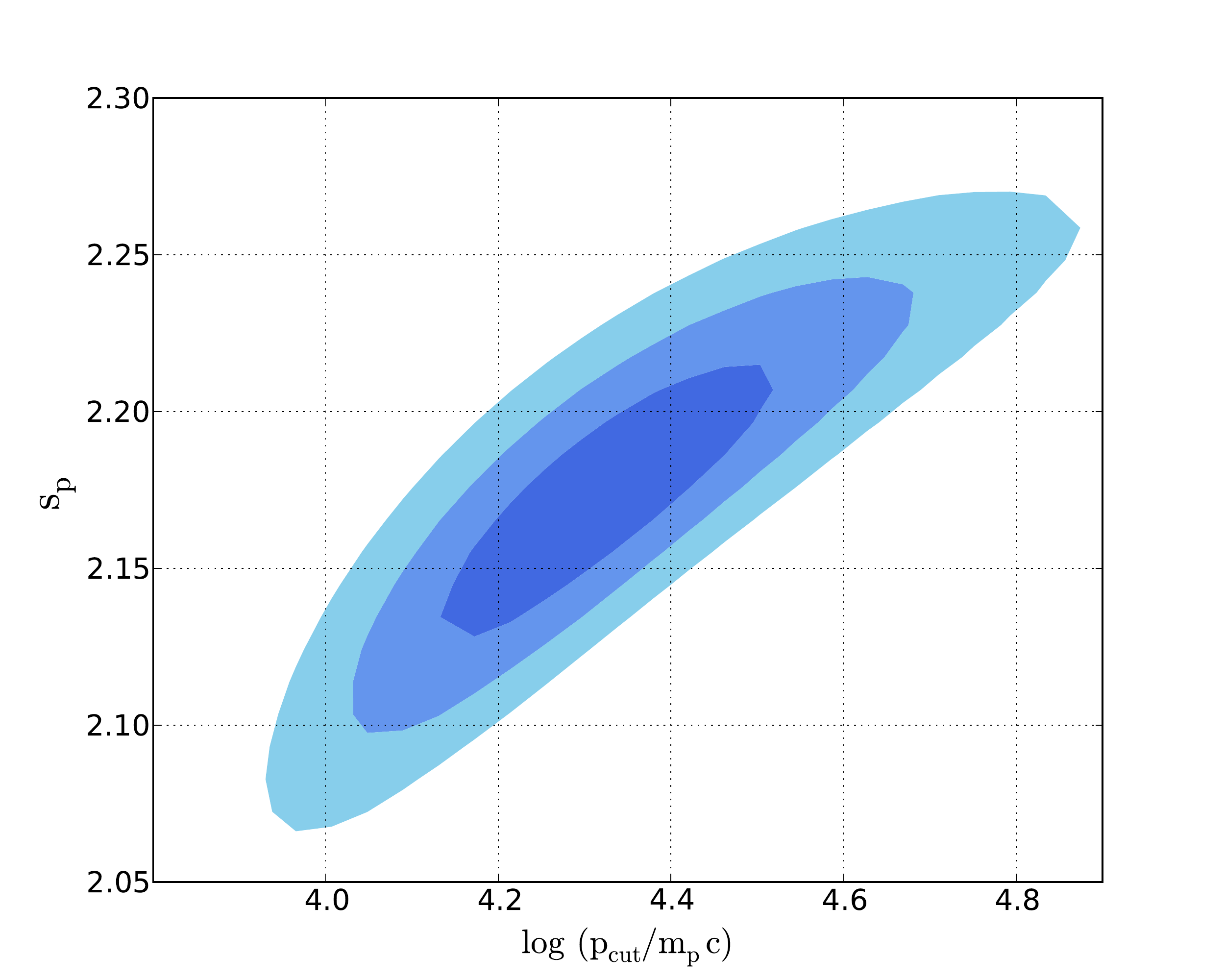}
\caption{\protect \small Purely hadronic model: The confidence regions for the spectral index, $s$, and cut-off momentum,  $p_{cut}$. The dark-blue area corresponds to 68.3\% probability, or 1$\sigma$, medium-blue to 95.5\%, or 2$\sigma$, and the light-blue field to 99.7\%, or 3$\sigma$, respectively.}
\label{3sigma}
\end{center}
\end{figure}

We start with a purely hadronic model of the $\gamma$-ray emission from Cas~A. Using Equation~\ref{eq:spectra} we find the best-fit for the joint \textit{Fermi}-LAT and VERITAS data points, shown in Figure~\ref{hadbestfit}. The corresponding best-fit parameters, with $\frac{\chi^2}{d.o.f.}=1.38$, are $s_p=2.17$ and $p_{cut}=2.1\times 10^4 \,\mathrm{m_p c}$ (equivalent to $E_{cut}\approx 17\,\mathrm{TeV}$). More instructive than the best-fit model are the confidence regions of the parameters, revealed by $\mathrm{\Delta \chi^2=\chi^2-\chi^2_{min}}$. Therefore, we scan the $s_p-p_{cut,p}$ parameter space  and calculate $\Delta \chi^2$ while optimizing $N_0$. The results are shown in Figure~\ref{3sigma}. Here the dark-blue area represents $\mathrm{\Delta \chi^2<2.30}$, medium-blue $\mathrm{\Delta \chi^2<6.18}$ and the light-blue $\mathrm{\Delta \chi^2<11.83}$, which corresponds to $1\sigma$, $2\sigma$ and $3\sigma$, respectively \citep{ConfLev}. As seen from Figure~\ref{3sigma}, the canonical solution from DSA theory ($s=2.0$) is excluded with $>99.7\%$ confidence. Thus, in the case of a hadronic origin, the $\gamma$-ray data mandate a proton spectral index ($s_p\approx2.1-2.2$) softer than predicted by the standard DSA theory ($s=2.0$) or nonlinear DSA ($s<2.0$ at $p\gg mc$) \citep{NLDSA}. Figure~\ref{3sigma} indicates a cut-off with $p_{cut,p}\sim 10^4\,\mathrm{m_p c}$, in full agreement with~\cite{Ahnen2017}, who concluded that Cas~A is not a PeVatron.

In the next step, we determine the electron spectrum for the global model of the broadband emission. The electron power-law index, $s_e\approx 2.5$, is entirely fixed by the radio data \citep{RadioData}, and the X-ray flux \citep{SuzakuData} is well explained by the synchrotron cut off.  A minor discrepancy occurs above 100 keV where the \textit{INTEGRAL} spectral data \citep{Wang2016} suggest a spectral hardening, which might reflect an asymmetric explosion \citep{Wang2016} and thus cannot be included in our modeling. An alternative explanation involves weakly relativistic electrons emitting NTB, as we discuss in Section~\ref{sec:lepto-had}.

\cite{Lee2014} found that the upstream gas density for Cas~A lies in the range 0.6 to 1.2~cm$^{-3}$. In this work we follow \cite{Lee2014} and use $n_{\mathrm{H}}= 1.0\,\mathrm{cm^{-3}}$ for simplicity. 
In order for the IC component not to dominate the $\gamma$-ray production from hadrons, the magnetic field in the downstream region needs to be at least $\sim$450~$\mu$G, and we use this minimum value in the model. This magnetic-field strength is compatible with the results of \cite{CasA_Zir&Ahar} and \cite{Sato_2018} who argued that for Cas~A, $B\sim 0.5-1 \, \mathrm{m G}$. For a magnetic field this strong ($\sim$450~$\mu$G) the thickness of the X-ray rims must reflect synchrotron energy losses of the radiating electrons~\citep{Parizot2006}.

\begin{figure}[htbp]
\begin{center}
\includegraphics[scale=0.7]{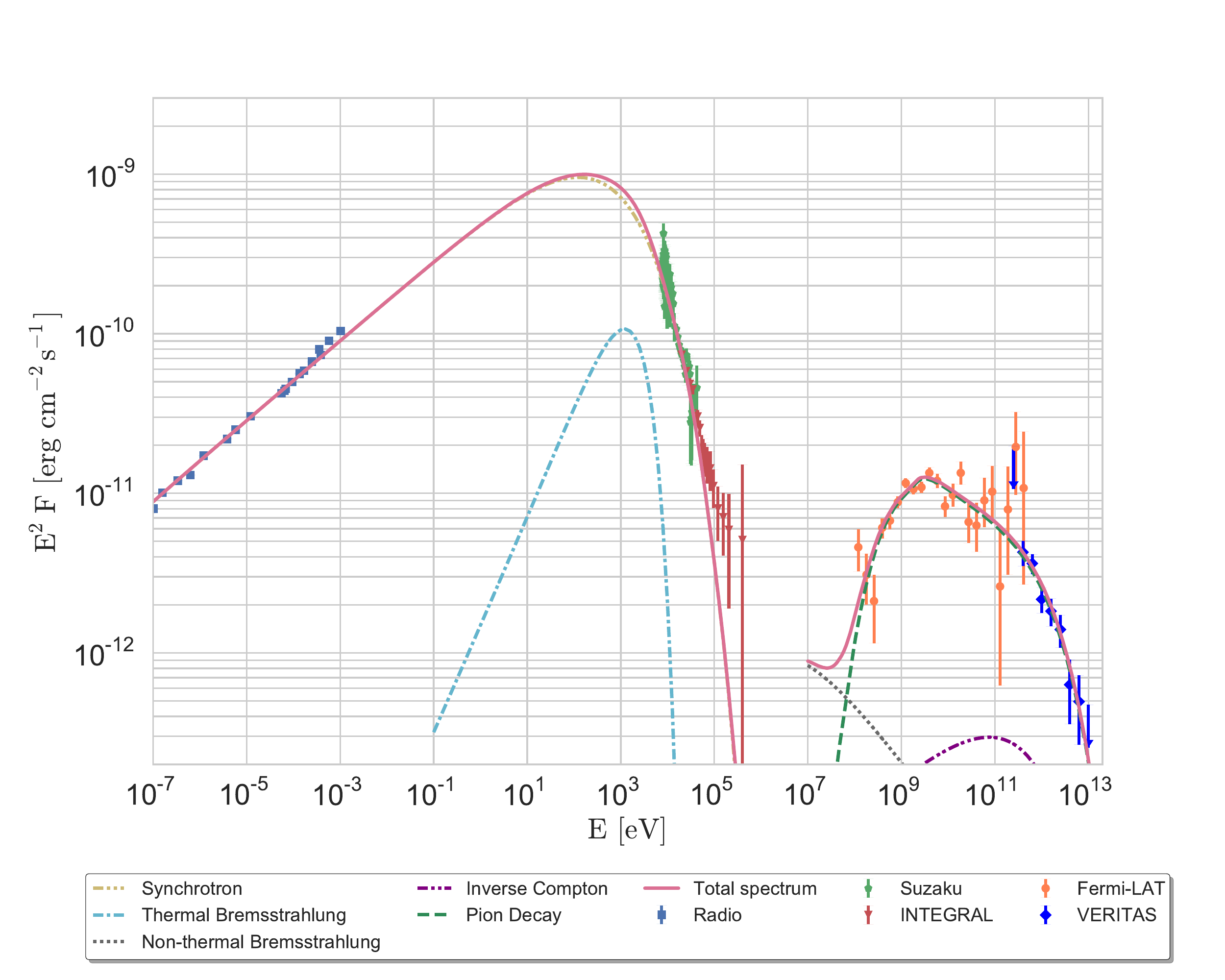}
\caption{\protect \small Model~I: Hadronic model with downstream magnetic field $B\approx 450\,\mathrm{\mu G}$ and upstream gas density $\mathrm{n_H}=1\,\mathrm{cm^{-3}}$. The radio data are taken from \cite{RadioData}; X-ray data from \cite{SuzakuData} and \cite{Wang2016}.}
\label{mod1}
\end{center}
\end{figure}

The entire SED is presented in Figure~\ref{mod1}, and the corresponding model parameters are summarized in Table~\ref{Theor_param} (Model~I). The hadronic component (green dashed line) is the best-fit spectrum presented in Figure~\ref{hadbestfit}. Besides the marginal IC contribution, we obtain a negligible NTB component, which we calculate starting from 10~MeV. While the spectral shape of the electrons for energies above $\sim$100~MeV can be constrained by the radio data, no data exist to test the spectral shape for electrons with energies below $\sim$100~MeV. Consequently, accurate modeling of the NTB radiation below $\sim$10~MeV, which corresponds to $\sim$100~MeV electron energy, is not possible. Therefore, in our modeling the total photon spectrum disconnects between 100~keV and 10~MeV. The electron temperature, $T_e$, is chosen according to \cite{SuzakuData}, and the thermal-bremsstrahlung emission provides a moderate contribution to the X-ray flux. The main reason for the rather insignificant thermal and NTB contributions is a relatively low plasma density in the downstream region
given for a strong shock by $n_{\mathrm{H, d}}=4 n_\mathrm{H}$.  

\begin{figure}[!phtb]
\begin{center}
\includegraphics[scale=0.6]{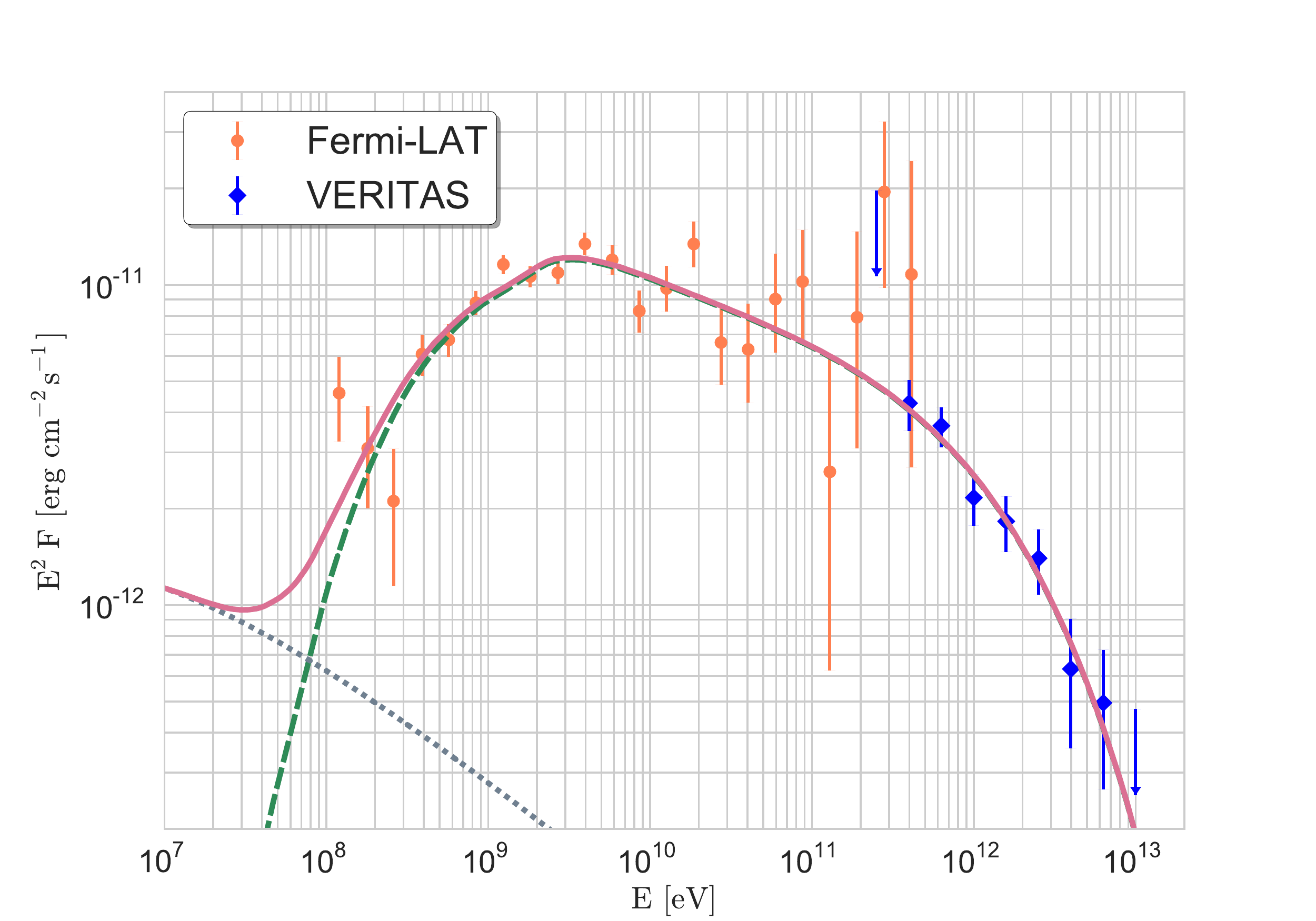}
\caption{\protect \small Best-fit for the hadronic component (green dashed line) plus non-thermal bremsstrahlung (blue dotted line); the total spectrum (pink solid line) with $\chi^2=35.50$ and $d.o.f.=25$ ($\chi^2/d.o.f.=1.42$).}
\label{pion_plus_ntb}
\end{center}
\end{figure}

Finally, we test if the increasing $\gamma$-ray flux at $\sim$100 MeV can be explained by NTB. Indeed, at first glance the two lowest-energy Fermi data points suggest the presence of an additional emission besides the pion bump, such as NTB. Performing the $\chi^2$-test after taking into account both NTB and neutral-pion decay, we find, however, that a negligible NTB contribution is preferred. The corresponding best-fit with $\frac{\chi^2}{d.o.f.}=1.42$ is presented in Figure~\ref{pion_plus_ntb}. Nevertheless, Cas~A has been considered for a long time as the best candidate for detecting NTB \citep{Cowsik, Allen_2008}. Therefore, we investigate the possibility of a lepto-hadronic model for the observed $\gamma$-ray spectrum of Cas~A in the following section. 

\subsection{Lepto-hadronic model}
\label{sec:lepto-had}
In this section we determine the observable limits on the presence of NTB and establish a model with a maximum possible NTB contribution.

In the framework of our one-zone model, NTB at a few hundred MeV is emitted by the same electrons that produce radio synchrotron emission at a few hundred MHz, and so a flux comparison between the radio data and the Fermi points at $\sim$100 MeV, ($F_{\mathrm{1\,GHz}}/F_{\mathrm{100\,MeV}}$), determines the relation between the average gas density and the minimum magnetic-field strength. Choosing the pre-shock gas density according to \cite{Lee2014}, $n_\mathrm{H}= 1.0\,\mathrm{cm^{-3}}$, we obtain for the minimum downstream magnetic-field strength: $B_\mathrm{min}\approx 150\,\mathrm{\mu G}$. Any weaker magnetic field would lead to NTB overshooting of the data points at $\sim$100~MeV. 

In general, the emission coefficients for synchrotron and NTB scale with magnetic-field strength and gas number density, respectively, as

\begin{equation}
    j_{sy}\propto B^{\frac{1+s_e}{2}}\qquad \qquad\textrm{and}\qquad \qquad j_{ntb}\propto n_\mathrm{H}\,.
\end{equation}
Therefore, to sustain constant synchrotron and NTB-flux ratio, the following condition for downstream magnetic field and ambient hydrogen number density has to be fulfilled:
\begin{equation}
    \left(\frac{B}{150\,\mathrm{\mu G}}\right)^{\frac{1+s_e}{2}}= \left(\frac{n_\mathrm{H}}{1\,\mathrm{cm}^{-3}}\right)\ .
     \label{MFvsD_eq}
\end{equation}
Aside from this case, the NTB component becomes suppressed with increasing magnetic field but constant gas density. Starting from some critical magnetic-field value the overall $\gamma$-emission becomes hadron dominated, as discussed in Section~\ref{sec:had_mod}. The minimum post-shock magnetic field for Cas~A is therefore given by
\begin{equation}
     B\gtrsim150\ \mathrm{\mu G}\,\left(\frac{n_\mathrm{H}}{\mathrm{cm}^{-3}}\right)^{\frac{2}{1+s_{e}}}\,,
     \label{MFvsD}
\end{equation}
as can be recognized from Equation~\ref{MFvsD_eq}. The minimum magnetic field deduced from potential NTB contribution depends on the ambient density of the remnant. The density uncertainties provided by \cite{Lee2014}, suggest that the minimum magnetic-field value may vary from $110\,\mathrm{\mu G}$ to  $170\,\mathrm{\mu G}$.

Having established the strength of the magnetic field inside Cas~A, we immediately find several consequences. First, given the age of the remnant, $\sim 10^{10}$~s, only electrons with Lorentz factors $\gamma\gg 10^6$ can be affected by energy losses. The resulting IC peak, which is calculated from a combination of CMB and FIR target-photon fields, would lie near $100$~GeV in the spectrum, and its spectral shape would be incompatible with that measured in the GeV band. The second consequence is that the peak energy flux of the IC component must be about a factor $U_\mathrm{mag}/(U_\mathrm{cmb}+U_\mathrm{fir})\simeq 250$ lower than that of the near-UV synchrotron emission radiated by the same electrons \citep{Pohl1996}. Consequently, the IC peak at 100 GeV is roughly a factor of 3 below the observed $\gamma$-ray flux and thus, IC emission alone can hardly provide the bulk of the $\gamma$-ray emission at $100$~GeV. It does contribute to a significant part of it though, and the highest-energy TeV emission is fully accounted for by the highest-energy IC contribution. Both points indicate that an additional radiation component, such as from neutral-pion decay, is required. Therefore, we conclude that a purely leptonic model is very unlikely.

\begin{figure}[htbp]
\begin{center}
\includegraphics[scale=0.7]{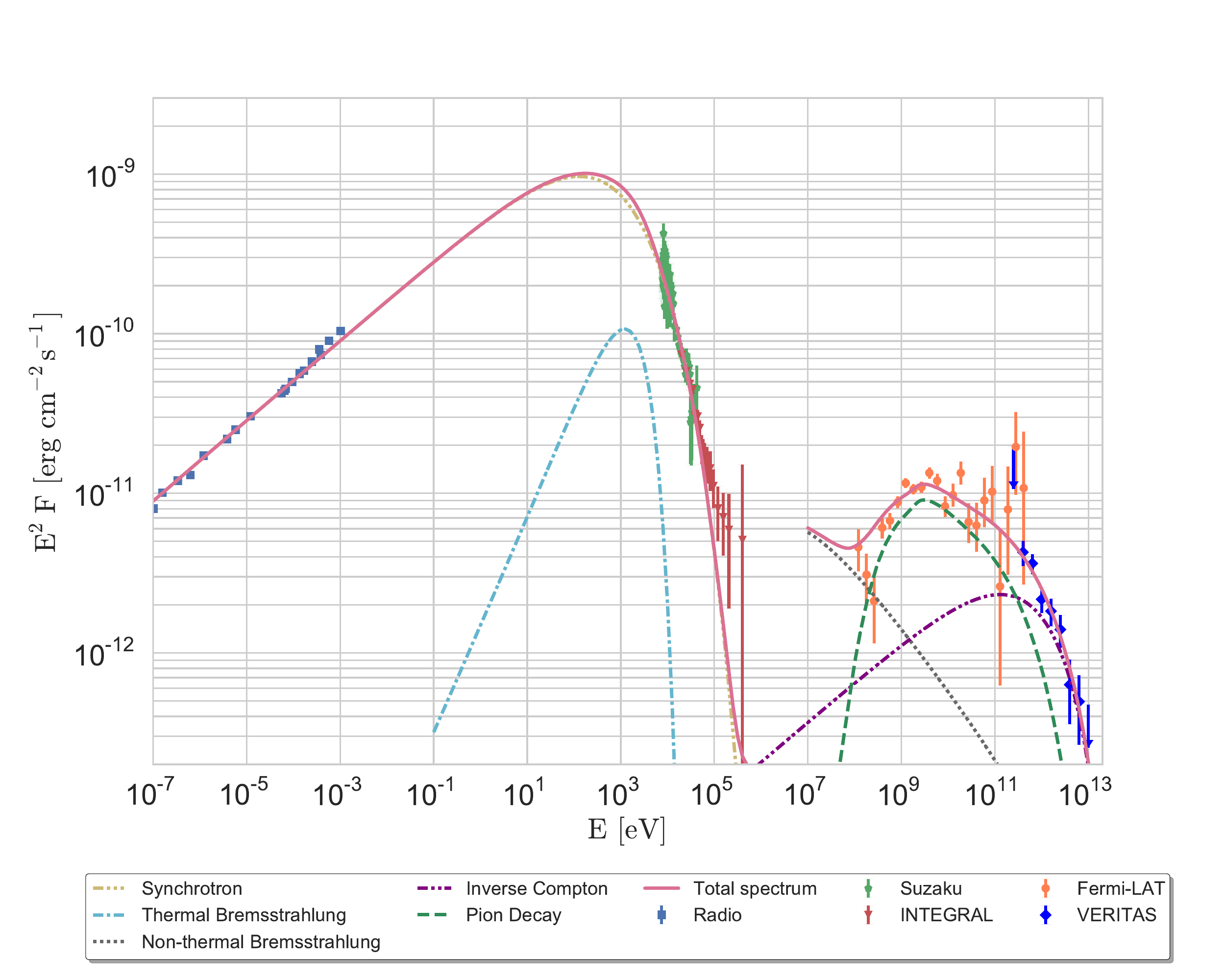}
\caption{\protect \small Model~II: Lepto-hadronic model with a minimum post-shock magnetic field $B\approx 150\,\mathrm{\mu G}$ and ambient gas density $\mathrm{n_H}=1\,\mathrm{cm^{-3}}$. The radio data are taken from \cite{RadioData}; X-ray data from \cite{SuzakuData} and \cite{Wang2016}.}
\label{mod2}
\end{center}
\end{figure}

The lepto-hadronic case (Model~II) with a maximum possible NTB component that is consistent with the Fermi data points is shown in Figure~\ref{mod2}. The IC peak (purple dash-dot-dotted line) located at $\sim$100~GeV sets an additional constraint on the magnetic field inside Cas~A. Decreasing the magnetic field would enhance the IC contribution, which would exceed the TeV-flux measured with VERITAS (blue diamond-shaped points in Figure~\ref{mod2}). Thus, both IC and NTB provide the same lower limit for the post-shock magnetic field,  $\sim 150\,\mathrm{\mu G}$. In contrast to NTB, IC does not scale with the gas density. Therefore, it provides an independent constraint on the magnetic-field value and implies that $B<150\,\mathrm{\mu G}$ is highly unlikely for Cas~A.

Despite a significant NTB contribution, $\gamma$-ray data in the GeV and higher MeV band are adequately explained by the pion bump and the discrimination between lepto-hadronic and purely hadronic models remains vague. Table~\ref{Theor_param} presents the parameters for the global lepto-hadronic model (Model~II). The normalization factor, $N_{0,e}$,  and the cut-off momentum of the electron spectrum, $p_{cut,e}$, are readjusted to fit the radio data for the weaker magnetic field. Since the cut off at TeV energies is largely reproduced by the IC, the proton spectrum cuts off already at roughly 6 TeV. Alternatively, the hadronic contribution at TeV energies can be reduced by assuming the proton spectral index softer than 2.17.

An advantage of the lepto-hadronic model is a possible explanation for the hardening of the X-ray spectrum above 100 keV observed with \textit{INTEGRAL} \citep{Wang2016} by emission from non-relativistic electrons radiating NTB. This idea is supported by the findings of \cite{Allen_2008}, who analyzed the X-ray data of Cas~A and concluded that, in the energy range 10-32~keV, NTB exceeds the synchrotron radiation by a factor 2 to 3. A logical extrapolation is that the non-relativistic electrons that are not in thermal equilibrium can provide a significant NTB contribution in the range 100~keV - 1~MeV and thus explain the hard X-ray spectrum. As mentioned above, we do not model this explicitly because we lack the exact shape of the electron spectrum at lower energies.


Note that, in contrast to the hadron-dominated model, in which we are able to use a chi-squared fit to the $>$100 MeV data, we follow a "fit by eye" process (as for example in \cite{Zhang:2019ycw}) for the lepto-hadronic scenario. The lepto-hadronic scenario includes the NTB and IC components and needs to incorporate the entire SED, making a formal fit and interpretation of the chi-squared from multiple instruments with very different statistical and systematic errors considerably more challenging.   We have also chosen a case with the minimum possible magnetic field inside Cas~A, which, as described above, provides the maximal (not best-fit) leptonic contributions.

\subsection{Discussion}
The observed radio spectrum  of Cas~A constrains the spectral index of the electrons to be $s_e\approx2.5$, and the $\gamma$-ray data favor a softer proton spectrum, $s_p\approx2.17$, than predicted by DSA. One possible explanation involves effects arising from turbulence growth and damping \citep{Malkov2011,Brose2016}. Alternatively, quasi-perpendicular shocks in young SNRs can steepen the spectral index \citep{Bell2011}. In the case of a young core-collapse SNR like Cas~A, the hydrodynamical structure of the progenitor wind zone and acceleration at the reverse shock can significantly modify the particle spectra \citep{Atoyan2000, IgorT_CC, CasA_Zir&Ahar}. The detection of X-ray synchrotron radiation in the interior of Cas~A suggests particle acceleration at the reverse shock \citep{Gotthelf2001,Uchiyama2008,RSxray}. However, newer data indicate that essentially all of the $>15 \U{keV}$ synchrotron flux is produced in small knots located in the 3D interior of the remnant, rather than a surface like the reverse shock \citep{Grefenstette2014}.  Finally, stochastic re-acceleration of electrons behind the forward shock may be able to soften the spectrum over 3 decades in synchrotron frequency \citep{reacc}. In the present work, we follow a simple procedure to address the most important conclusions: determination of the minimum magnetic field strength; confirmation of the pion bump and the corresponding proton cut-off energy. More sophisticated models (including, e.g., asymmetric explosion, time-dependent hydrodynamic simulations, acceleration at the reverse shock, magnetic turbulence and stochastic re-acceleration of particles) are needed, to further differentiate between competing scenarios concerning particle acceleration in SNRs.

The total cosmic ray energy for the hadron-dominated (Model~I) and lepto-hadronic (Model~II) models considered here is found to be $E_\mathrm{CR}\approx1.7\times 10^{50}$~erg and $E_\mathrm{CR}\approx1.2\times10^{50}$~erg, respectively. These numbers roughly represent the total energy that went into the particles as they accumulated over the entire evolution time of the remnant. Unfortunately, there is no easy way to ascertain the original explosion energy of Cas~A, $E_\mathrm{SN}$: the estimations vary between 2$\times10^{51}$~erg and 5$\times10^{51}$~erg \citep{LamingHwang, Chevalier2003, Schure_2008, Lee2014, Orlando_2016}. This suggests that the fraction of the explosion energy expended in accelerating particles is between 2\% and 9\%. Being a very young SNR, Cas~A is very likely in the ejecta-dominated phase~\citep{Morse_2004}, implying that only a fraction of its explosion energy can be currently extracted from the shock. The full energy becomes available after the SNR enters the Sedov-Taylor stage. In that case, the above numbers may not indicate accurately the acceleration efficiency of the remnant. \cite{Truelove_1999} suggested that Cas~A is in transition from the ejecta-dominated to the Sedov-Taylor stage. To verify this, we follow calculations in~\cite{Dwark_2013}, who assumed that Cas~A is still in the free-expansion phase and expands into a wind with density profile $\rho \propto r^{-2}$. The maximum shock energy that is available for particle acceleration is found to be:
\begin{equation}
  E_\mathrm{acc}=\frac{2\pi m^3}{(3m-2)}\,\frac{\rho_\mathrm{u} R_\mathrm{sh}^5} {t_\mathrm{age}^{2}} \qquad \textrm{with} \qquad m=\frac{(n-3)}{(n-2)}\,.
  \label{eq:cr_energy}
\end{equation}
Here $\rho_\mathrm{u}$ is pre-shock gas density, $R_\mathrm{sh}$ is shock radius and $t_\mathrm{age}$ is age of the remnant. The expansion parameter, defined as $m=d\ln{R_\mathrm{sh}}/{d\ln{t}}$, is fixed by the ejecta-density profile, $\rho_\mathrm{ej}\propto r^{-n}$, with $n>5$ \citep[e.g.][]{Chevalier_1982}. A reasonable value for $n$ is given by \cite{Matzner_1999} who find that a red supergiant star with a radiative envelope has $n\approx10$. Assuming this ejecta profile and taking typical values for Cas~A: $R_\mathrm{sh}=2.5$~pc, $\rho_\mathrm{u}=2.34\times 10^{-24}\,\mathrm{g\,cm^{-3}}$ and $t_\mathrm{age}=350$~years, we obtain the maximum shock energy available at $E_\mathrm{acc} \approx 3.5\times 10^{51} \, \mathrm{erg}$.
This result shows that the maximum energy available for particle acceleration in the ejecta-dominated phase is of the same order as the total explosion energy of Cas~A, $E_\mathrm{SN}\approx 2\times10^{51}- 5\times10^{51}$~erg, that is presented in literature \citep{LamingHwang, Chevalier2003, Schure_2008, Lee2014, Orlando_2016}. This indicates that a large fraction of the explosion energy is available at the shock front. Therefore, Cas~A is not far from the Sedov-Taylor stage. Our estimation of $2\%-9\%$ of explosion energy is thus appropriate. Further, $E_\mathrm{acc} \approx 3.5\times10^{51} \, \mathrm{erg}$ implies that the acceleration efficiency (defined as $\eta=E_\mathrm{CR}/E_\mathrm{acc}$) is $\eta\approx0.05$ and $\eta\approx0.03$ for hadronic and lepto-hadronic scenarios, respectively. However, one should treat these conclusions with caution, since the values we used for the parameters in Equation~\ref{eq:cr_energy} are not precisely known. Our result is consistent with the total cosmic ray energy $\sim 9.9\times 10^{49}$ erg presented by the MAGIC collaboration \citep{Ahnen2017} and exceeds the value $\sim 4\times 10^{49}$ erg found using \textit{Fermi}-LAT \citep{Yuan2013}.

We find that IC and NTB obviously cannot account for the emission around 10~GeV, and thus a hadronic component is clearly needed. The maximum energies obtained for protons are 21~TeV and 6~TeV for the purely hadronic and lepto-hadronic models, respectively. These values are similar to the previous results of \cite{Yuan2013} (10~TeV) and \cite{Ahnen2017}(12~TeV).

\section{Conclusions}
In this work, we have presented a deep study of the supernova remnant Cas~A using 10.8 years of \textit{Fermi}-LAT and 65 hours of VERITAS data. The centroid positions from \textit{Fermi}-LAT and VERITAS measurements are found to be consistent, within errors, and lie inside the remnant. Since the size of the remnant is comparable to the PSF of the \textit{Fermi}-LAT and VERITAS instruments, it is difficult to determine whether the emission is coming from the forward or the reverse shock within the SNR. More sensitive instruments, in the future, such as the Cherenkov Telescope Array (CTA) \citep{CTA2013}, will allow us to perform better measurements on the morphology of this source. Above $100 \U{MeV}$, a spectral index change from $1.3$ to $2.1$ is measured at an energy of $1.3 \pm 0.4 \U{GeV}$ with the \textit{Fermi}-LAT data, which is consistent with previous observations \citep{Yuan2013} and can be explained by $\gamma$-ray emission produced through neutral-pion decay. In addition, a joint spectral fit of \textit{Fermi}-LAT and VERITAS spectral data from $\sim 2 \U{GeV}$ to $10 \U{TeV}$ prefers an exponential cut-off power-law to a single power-law model. The cut-off energy found using \textit{Fermi}-LAT and VERITAS data is estimated to be $2.3 \pm 0.5 \U{TeV}$. This is compatible with the cut-off energy found by the MAGIC collaboration using only MAGIC data \citep{Ahnen2017}. This shows that the Cas~A SNR is unlikely to be a source of PeV cosmic rays.

In the theoretical part of this work we took radio \citep{RadioData} and X-ray \citep{SuzakuData, Wang2016} observations into account. Considering the entire multi-wavelength spectrum of Cas~A, we used a global one-zone model assuming power-law particle spectra with an exponential cut off. Two different scenarios, a hadron-dominated case (Model~I) and a lepto-hadronic model (Model~II) are presented. Furthermore, in agreement with previous studies on the SED of Cas~A \citep{Araya_2010, Saha_2014}; a
purely leptonic model is excluded under the assumption of a one-zone scenario, leading to the conclusion that proton acceleration up to TeV energies
is clearly evident. The resulting pion bump reflects a slightly softer spectral index
for the proton spectrum, $s_p\approx 2.17$, than the canonical DSA predictions (both linear and non-linear versions~\citep{NLDSA}). We exclude the canonical DSA solution of $s=2.0$ with $3\sigma$ confidence. The total energy converted into cosmic rays is at least $10^{50}$~erg, giving an acceleration efficiency $\eta\approx0.03-0.05$.

Although Cas~A is the best SNR candidate for NTB emission  \citep{Cowsik, Allen_2008}, our observations
do not indicate any evidence for a NTB flux above 100 MeV. A clear determination may be achieved with the photon measurements extended down to the MeV energy range. Future experiments, such as AMEGO\footnote{\url{https://asd.gsfc.nasa.gov/amego/}} (All-Sky Medium Energy Gamma-ray Observatory), may shed light on that issue. Nevertheless, assuming a potential
NTB presence in Cas~A, we set a minimum value for the magnetic-field strength inside the remnant $B_\mathrm{min}\approx 150\, \mathrm{\mu G}$. This value is independently confirmed by
the IC peak. Therefore, it is clear that the magnetic field inside the Cas~A SNR is efficiently amplified, when compared to the interstellar-medium field.

%
\acknowledgements{This research is supported by grants from the U.S. Department of Energy Office of Science, the U.S. National Science Foundation and the Smithsonian Institution, and by NSERC in Canada. We acknowledge the excellent work of the technical support staff at the Fred Lawrence Whipple Observatory and at the collaborating institutions in the construction and operation of the instrument. VVD's work is supported by NSF grant 1911061 awarded to the University of Chicago (PI: Vikram Dwarkadas).}
\bibliography{listof_cited_papers}

\begin{thebibliography}{}
\expandafter\ifx\csname natexlab\endcsname\relax\def\natexlab#1{#1}\fi

\bibitem[{{Abdo} {et~al.}(2009)}]{Abdo2009a}
{Abdo}, A.~A., {et~al.} 2009, \apjl, 706, L1

\bibitem[{{Abdo} {et~al.}(2010)}]{Abdo2010}
{Abdo}, A.~A., {et~al.} 2010, \apjl, 710, L92

\bibitem[{{Acciari} {et~al.}(2008)}]{Acciari2008}
{Acciari}, V.~A., {et~al.} 2008, \apj, 679, 1427

\bibitem[{{Acciari} {et~al.}(2010)}]{Acciari2010}
{Acciari}, V.~A., {et~al.} 2010, \apj, 714, 163

\bibitem[{{Acharya} {et~al.}(2013){Acharya}, {Actis}, {Aghajani}, {Agnetta},
  {Aguilar}, {Aharonian}, {Ajello}, {Akhperjanian}, {Alcubierre},
  {Aleksi{\'c}}, \& et~al.}]{CTA2013}
{Acharya}, B.~S., {Actis}, M., {Aghajani}, T., {et~al.} 2013, Astroparticle
  Physics, 43, 3

\bibitem[{{Ackermann} {et~al.}(2013)}]{Ackermann2013}
{Ackermann}, M., {et~al.} 2013, Science, 339, 807

\bibitem[{{Aharonian} {et~al.}(2001)}]{Aharonian2001}
{Aharonian}, F., {et~al.} 2001, \aap, 370, 112

\bibitem[{{Ahnen} {et~al.}(2017)}]{Ahnen2017}
{Ahnen}, M.~L., {et~al.} 2017, \mnras, 472, 2956

\bibitem[{{Albert} {et~al.}(2007)}]{Albert2007}
{Albert}, J., {et~al.} 2007, \aap, 474, 937

\bibitem[{{Allen} {et~al.}(2008){Allen}, {Stage}, \& {Houck}}]{Allen_2008}
{Allen}, G.~E., {Stage}, M.~D., \& {Houck}, J.~C. 2008, International Cosmic
  Ray Conference, 2, 839

\bibitem[{{Araya} \& {Cui}(2010)}]{Araya_2010}
{Araya}, M., \& {Cui}, W. 2010, \apj, 720, 20

\bibitem[{{Atoyan} {et~al.}(2000){Atoyan}, {Aharonian}, {Tuffs}, \&
  {V{\"o}lk}}]{Atoyan2000}
{Atoyan}, A.~M., {Aharonian}, F.~A., {Tuffs}, R.~J., \& {V{\"o}lk}, H.~J. 2000,
  \aap, 355, 211

\bibitem[{{Atwood} {et~al.}(2013){Atwood}, {Albert}, {Baldini}, {Tinivella},
  {Bregeon}, {Pesce-Rollins}, {Sgr{\`o}}, {Bruel}, {Charles}, {Drlica-Wagner},
  {Franckowiak}, {Jogler}, {Rochester}, {Usher}, {Wood}, {Cohen-Tanugi}, \&
  {S.~Zimmer for the Fermi-LAT Collaboration}}]{Atwood2013}
{Atwood}, W., {Albert}, A., {Baldini}, L., {et~al.} 2013, ArXiv e-prints,
  arXiv:1303.3514

\bibitem[{{Atwood} {et~al.}(2009){Atwood}, {Abdo}, {Ackermann}, {Althouse},
  {Anderson}, {Axelsson}, {Baldini}, {Ballet}, {Band}, {Barbiellini}, \&
  et~al.}]{Atwood2009}
{Atwood}, W.~B., {Abdo}, A.~A., {Ackermann}, M., {et~al.} 2009, \apj, 697, 1071

\bibitem[{{Axford} {et~al.}(1977){Axford}, {Leer}, \& {Skadron}}]{Axford1977}
{Axford}, W.~I., {Leer}, E., \& {Skadron}, G. 1977, International Cosmic Ray
  Conference, 11, 132

\bibitem[{{Baade} \& {Zwicky}(1934)}]{Baade1934}
{Baade}, W., \& {Zwicky}, F. 1934, Physical Review, 46, 76

\bibitem[{{Baars} {et~al.}(1977){Baars}, {Genzel}, {Pauliny-Toth}, \&
  {Witzel}}]{Baars1977}
{Baars}, J.~W.~M., {Genzel}, R., {Pauliny-Toth}, I.~I.~K., \& {Witzel}, A.
  1977, \aap, 61, 99

\bibitem[{{Bell}(1978{\natexlab{a}})}]{Bell1978I}
{Bell}, A.~R. 1978{\natexlab{a}}, \mnras, 182, 147

\bibitem[{{Bell}(1978{\natexlab{b}})}]{Bell1978II}
{Bell}, A.~R. 1978{\natexlab{b}}, \mnras, 182, 443

\bibitem[{{Bell} {et~al.}(1975){Bell}, {Gull}, \& {Kenderdine}}]{Bell1975}
{Bell}, A.~R., {Gull}, S.~F., \& {Kenderdine}, S. 1975, \nat, 257, 463

\bibitem[{{Bell} {et~al.}(2011){Bell}, {Schure}, \& {Reville}}]{Bell2011}
{Bell}, A.~R., {Schure}, K.~M., \& {Reville}, B. 2011, \mnras, 418, 1208

\bibitem[{{Berge} {et~al.}(2007){Berge}, {Funk}, \& {Hinton}}]{Berge2007}
{Berge}, D., {Funk}, S., \& {Hinton}, J. 2007, A\&A, 466, 1219

\bibitem[{{Blandford} \& {Ostriker}(1978)}]{Blandford1978}
{Blandford}, R.~D., \& {Ostriker}, J.~P. 1978, \apjl, 221, L29

\bibitem[{Blumenthal \& Gould(1970)}]{synchrotron&IC}
Blumenthal, G.~R., \& Gould, R.~J. 1970, Rev. Mod. Phys., 42, 237

\bibitem[{{Braun} {et~al.}(1987){Braun}, {Gull}, \& {Perley}}]{Braun1987}
{Braun}, R., {Gull}, S.~F., \& {Perley}, R.~A. 1987, \nat, 327, 395

\bibitem[{{Brose} {et~al.}(2016){Brose}, {Telezhinsky}, \& {Pohl}}]{Brose2016}
{Brose}, R., {Telezhinsky}, I., \& {Pohl}, M. 2016, \aap, 593, A20

\bibitem[{{Chevalier}(1982)}]{Chevalier_1982}
{Chevalier}, R.~A. 1982, \apj, 258, 790

\bibitem[{{Chevalier} \& {Oishi}(2003)}]{Chevalier2003}
{Chevalier}, R.~A., \& {Oishi}, J. 2003, \apjl, 593, L23

\bibitem[{{Cogan}(2008)}]{Cogan2008}
{Cogan}, P. 2008, International Cosmic Ray Conference, 3, 1385

\bibitem[{Cowsik \& Sarkar(1980)}]{Cowsik}
Cowsik, R., \& Sarkar, S. 1980, Mon. Not. Roy. Astron. Soc., 191, 855

\bibitem[{Degrange \& Fontaine(2015)}]{DEGRANGE2015}
Degrange, B., \& Fontaine, G. 2015, Comptes Rendus Physique, 16, 587 ,
  gamma-ray astronomy / Astronomie des rayons gamma

\bibitem[{{DeLaney} {et~al.}(2014){DeLaney}, {Kassim}, {Rudnick}, \&
  {Perley}}]{Delaney2014}
{DeLaney}, T., {Kassim}, N.~E., {Rudnick}, L., \& {Perley}, R.~A. 2014, \apj,
  785, 7

\bibitem[{{Dwarkadas}(2013)}]{Dwark_2013}
{Dwarkadas}, V.~V. 2013, \mnras, 434, 3368

\bibitem[{{Fesen} {et~al.}(2006){Fesen}, {Hammell}, {Morse}, {Chevalier},
  {Borkowski}, {Dopita}, {Gerardy}, {Lawrence}, {Raymond}, \& {van den
  Bergh}}]{Fesen2006}
{Fesen}, R.~A., {Hammell}, M.~C., {Morse}, J., {et~al.} 2006, \apj, 645, 283

\bibitem[{{Fomin} {et~al.}(1994){Fomin}, {Stepanian}, {Lamb}, {Lewis}, {Punch},
  \& {Weekes}}]{Fomin1994}
{Fomin}, V.~P., {Stepanian}, A.~A., {Lamb}, R.~C., {et~al.} 1994, Astroparticle
  Physics, 2, 137

\bibitem[{{Freeman} {et~al.}(2001){Freeman}, {Doe}, \&
  {Siemiginowska}}]{Sherpa2001}
{Freeman}, P., {Doe}, S., \& {Siemiginowska}, A. 2001, in \procspie, Vol. 4477,
  Astronomical Data Analysis, ed. J.-L. {Starck} \& F.~D. {Murtagh}, 76--87

\bibitem[{{Gerardy} \& {Fesen}(2001)}]{Gerardy2001}
{Gerardy}, C.~L., \& {Fesen}, R.~A. 2001, \aj, 121, 2781

\bibitem[{Ginzburg \& Syrovatski{\u{\i}}(1966)}]{Ginzburg_1966}
Ginzburg, V.~L., \& Syrovatski{\u{\i}}, S.~I. 1966, Soviet Physics Uspekhi, 9,
  223

\bibitem[{{Gotthelf} {et~al.}(2001){Gotthelf}, {Koralesky}, {Rudnick}, {Jones},
  {Hwang}, \& {Petre}}]{Gotthelf2001}
{Gotthelf}, E.~V., {Koralesky}, B., {Rudnick}, L., {et~al.} 2001, \apjl, 552,
  L39

\bibitem[{{Grefenstette} {et~al.}(2015)}]{Grefenstette2014}
{Grefenstette}, B.~W., {et~al.} 2015, ApJ, 802, 15

\bibitem[{{Helder} \& {Vink}(2008)}]{RSxray}
{Helder}, E.~A., \& {Vink}, J. 2008, \apj, 686, 1094

\bibitem[{Hnatyk \& Petruk(1999)}]{thermal}
Hnatyk, B., \& Petruk, O. 1999, Astron. Astrophys., 344, 295

\bibitem[{{Holder} {et~al.}(2006)}]{Holder2006}
{Holder}, J., {et~al.} 2006, Astroparticle Physics, 25, 391

\bibitem[{{Holt} {et~al.}(1994){Holt}, {Gotthelf}, {Tsunemi}, \&
  {Negoro}}]{Holt1994}
{Holt}, S.~S., {Gotthelf}, E.~V., {Tsunemi}, H., \& {Negoro}, H. 1994, \pasj,
  46, L151

\bibitem[{{Huang} {et~al.}(2007){Huang}, {Park}, {Pohl}, \&
  {Daniels}}]{gammaProt}
{Huang}, C.-Y., {Park}, S.-E., {Pohl}, M., \& {Daniels}, C.~D. 2007,
  Astroparticle Physics, 27, 429

\bibitem[{Hwang {et~al.}(2004)Hwang, Laming, Badenes, Berendse, Blondin,
  Cioffi, DeLaney, Dewey, Fesen, Flanagan, Fryer, Ghavamian, Hughes, Morse,
  Plucinsky, Petre, Pohl, Rudnick, Sankrit, Slane, Smith, Vink, \&
  Warren}]{Hwang2004}
Hwang, U., Laming, J.~M., Badenes, C., {et~al.} 2004, The Astrophysical Journal
  Letters, 615, L117

\bibitem[{{Jones} {et~al.}(2003){Jones}, {Rudnick}, {DeLaney}, \&
  {Bowden}}]{Jones2003}
{Jones}, T.~J., {Rudnick}, L., {DeLaney}, T., \& {Bowden}, J. 2003, \apj, 587,
  227

\bibitem[{{Kassim} {et~al.}(1995){Kassim}, {Perley}, {Dwarakanath}, \&
  {Erickson}}]{Kassim1995}
{Kassim}, N.~E., {Perley}, R.~A., {Dwarakanath}, K.~S., \& {Erickson}, W.~C.
  1995, \apjl, 455, L59

\bibitem[{{Kieda} {et~al.}(2013)}]{Kieda2013}
{Kieda}, D., {et~al.} 2013, ArXiv e-prints, arXiv:1308.4849

\bibitem[{{Krause} {et~al.}(2008){Krause}, {Birkmann}, {Usuda}, {Hattori},
  {Goto}, {Rieke}, \& {Misselt}}]{Krause2008}
{Krause}, O., {Birkmann}, S.~M., {Usuda}, T., {et~al.} 2008, Science, 320, 1195

\bibitem[{{Krymskii}(1977)}]{Krymskii1977}
{Krymskii}, G.~F. 1977, Akademiia Nauk SSSR Doklady, 234, 1306

\bibitem[{{Kumar} {et~al.}(2015)}]{KumarICRC2015}
{Kumar}, S., {et~al.} 2015, ArXiv e-prints, arXiv:1508.07453

\bibitem[{Laming \& Hwang(2003)}]{LamingHwang}
Laming, J.~M., \& Hwang, U. 2003, The Astrophysical Journal, 597, 347

\bibitem[{{Lampton} {et~al.}(1976){Lampton}, {Margon}, \& {Bowyer}}]{ConfLev}
{Lampton}, M., {Margon}, B., \& {Bowyer}, S. 1976, \apj, 208, 177

\bibitem[{{Lee} {et~al.}(2014){Lee}, {Park}, {Hughes}, \& {Slane}}]{Lee2014}
{Lee}, J.-J., {Park}, S., {Hughes}, J.~P., \& {Slane}, P.~O. 2014, \apj, 789, 7

\bibitem[{{Li} \& {Ma}(1983)}]{LiMa1983}
{Li}, T.-P., \& {Ma}, Y.-Q. 1983, \apj, 272, 317

\bibitem[{Madhavan(2013)}]{MadhavanThesis2013}
Madhavan, A. 2013, PhD thesis, Iowa State University

\bibitem[{Maeda {et~al.}(2009)}]{SuzakuData}
Maeda, Y., {et~al.} 2009, Publ. Astron. Soc. Jap., 61, 1217

\bibitem[{{Maier} \& {Holder}(2017)}]{MaierED2017}
{Maier}, G., \& {Holder}, J. 2017, ArXiv e-prints, arXiv:1708.04048

\bibitem[{{Malkov} {et~al.}(2011){Malkov}, {Diamond}, \&
  {Sagdeev}}]{Malkov2011}
{Malkov}, M.~A., {Diamond}, P.~H., \& {Sagdeev}, R.~Z. 2011, Nature
  Communications, 2, 194

\bibitem[{{Malkov} \& {Drury}(2001)}]{NLDSA}
{Malkov}, M.~A., \& {Drury}, L.~O. 2001, Reports on Progress in Physics, 64,
  429

\bibitem[{{Matzner} \& {McKee}(1999)}]{Matzner_1999}
{Matzner}, C.~D., \& {McKee}, C.~F. 1999, \apj, 510, 379

\bibitem[{{Mezger} {et~al.}(1986){Mezger}, {Tuffs}, {Chini}, {Kreysa}, \&
  {Gemuend}}]{Mezger}
{Mezger}, P.~G., {Tuffs}, R.~J., {Chini}, R., {Kreysa}, E., \& {Gemuend}, H.-P.
  1986, \aap, 167, 145

\bibitem[{{Morse} {et~al.}(2004){Morse}, {Fesen}, {Chevalier}, {Borkowski},
  {Gerardy}, {Lawrence}, \& {van den Bergh}}]{Morse_2004}
{Morse}, J.~A., {Fesen}, R.~A., {Chevalier}, R.~A., {et~al.} 2004, \apj, 614,
  727

\bibitem[{{Orlando} {et~al.}(2016){Orlando}, {Miceli}, {Pumo}, \&
  {Bocchino}}]{Orlando_2016}
{Orlando}, S., {Miceli}, M., {Pumo}, M.~L., \& {Bocchino}, F. 2016, \apj, 822,
  22

\bibitem[{{Otte} {et~al.}(2011)}]{Nepomuk2011}
{Otte}, A., {et~al.} 2011, ArXiv e-prints, arXiv:1110.4702

\bibitem[{{Parizot} {et~al.}(2006){Parizot}, {Marcowith}, {Ballet}, \&
  {Gallant}}]{Parizot2006}
{Parizot}, E., {Marcowith}, A., {Ballet}, J., \& {Gallant}, Y.~A. 2006, \aap,
  453, 387

\bibitem[{{Park} {et~al.}(2015)}]{NaheeICRC2015}
{Park}, N., {et~al.} 2015, in International Cosmic Ray Conference, Vol.~34,
  34th International Cosmic Ray Conference (ICRC2015), 771

\bibitem[{{Perkins} {et~al.}(2009){Perkins}, {Maier}, \& {The VERITAS
  Collaboration}}]{Perkins2009}
{Perkins}, J.~S., {Maier}, G., \& {The VERITAS Collaboration}. 2009, ArXiv
  e-prints, arXiv:0912.3841

\bibitem[{{Pohl}(1996)}]{Pohl1996}
{Pohl}, M. 1996, \aap, 307, L57

\bibitem[{Pohl {et~al.}(2015)Pohl, Wilhelm, \& Telezhinsky}]{reacc}
Pohl, M., Wilhelm, A., \& Telezhinsky, I. 2015, Astron. Astrophys., 574, A43

\bibitem[{{Reed} {et~al.}(1995){Reed}, {Hester}, {Fabian}, \&
  {Winkler}}]{Reed1995}
{Reed}, J.~E., {Hester}, J.~J., {Fabian}, A.~C., \& {Winkler}, P.~F. 1995,
  \apj, 440, 706

\bibitem[{{Rho} {et~al.}(2003){Rho}, {Reynolds}, {Reach}, {Jarrett}, {Allen},
  \& {Wilson}}]{Rho2003}
{Rho}, J., {Reynolds}, S.~P., {Reach}, W.~T., {et~al.} 2003, \apj, 592, 299

\bibitem[{{Saha} {et~al.}(2014){Saha}, {Ergin}, {Majumdar}, {Bozkurt}, \&
  {Ercan}}]{Saha_2014}
{Saha}, L., {Ergin}, T., {Majumdar}, P., {Bozkurt}, M., \& {Ercan}, E.~N. 2014,
  \aap, 563, A88

\bibitem[{Sato {et~al.}(2018)Sato, Katsuda, Morii, Bamba, Hughes, Maeda,
  Ishida, \& Fraschetti}]{Sato_2018}
Sato, T., Katsuda, S., Morii, M., {et~al.} 2018, The Astrophysical Journal,
  853, 46

\bibitem[{{Schure} {et~al.}(2008){Schure}, {Vink}, {Garc{\'{\i}}a-Segura}, \&
  {Achterberg}}]{Schure_2008}
{Schure}, K.~M., {Vink}, J., {Garc{\'{\i}}a-Segura}, G., \& {Achterberg}, A.
  2008, \apj, 686, 399

\bibitem[{{Sommers} \& {Elbert}(1987)}]{sommers1986}
{Sommers}, P., \& {Elbert}, J.~W. 1987, Journal of Physics G Nuclear Physics,
  13, 553

\bibitem[{{Telezhinsky} {et~al.}(2013){Telezhinsky}, {Dwarkadas}, \&
  {Pohl}}]{IgorT_CC}
{Telezhinsky}, I., {Dwarkadas}, V.~V., \& {Pohl}, M. 2013, \aap, 552, A102

\bibitem[{{The Fermi-LAT collaboration}(2019)}]{4FGL2019}
{The Fermi-LAT collaboration}. 2019, arXiv e-prints, arXiv:1902.10045

\bibitem[{{Truelove} \& {McKee}(1999)}]{Truelove_1999}
{Truelove}, J.~K., \& {McKee}, C.~F. 1999, \apjs, 120, 299

\bibitem[{{Uchiyama} \& {Aharonian}(2008)}]{Uchiyama2008}
{Uchiyama}, Y., \& {Aharonian}, F.~A. 2008, \apjl, 677, L105

\bibitem[{{Vink} \& {Laming}(2003)}]{Vink_MF}
{Vink}, J., \& {Laming}, J.~M. 2003, \apj, 584, 758

\bibitem[{{Vinyaikin}(2014)}]{RadioData}
{Vinyaikin}, E.~N. 2014, Astronomy Reports, 58, 626

\bibitem[{{Wang} \& {Li}(2016)}]{Wang2016}
{Wang}, W., \& {Li}, Z. 2016, \apj, 825, 102

\bibitem[{Weekes {et~al.}(2002)}]{Weekes2002}
Weekes, T.~C., {et~al.} 2002, Astropart. Phys., 17, 221

\bibitem[{{Wood} {et~al.}(2017){Wood}, {Caputo}, {Charles}, {Di Mauro},
  {Magill}, \& {Jeremy Perkins for the Fermi-LAT
  Collaboration}}]{WoodFermipy2017}
{Wood}, M., {Caputo}, R., {Charles}, E., {et~al.} 2017, ArXiv e-prints,
  arXiv:1707.09551

\bibitem[{{Yuan} {et~al.}(2013){Yuan}, {Funk}, {J{\'o}hannesson}, {Lande},
  {Tibaldo}, \& {Uchiyama}}]{Yuan2013}
{Yuan}, Y., {Funk}, S., {J{\'o}hannesson}, G., {et~al.} 2013, \apj, 779, 117

\bibitem[{{Zhang} \& {Liu}(2019)}]{2019ApJ...874...98Z}
{Zhang}, X., \& {Liu}, S. 2019, \apj, 874, 98

\bibitem[{Zhang \& Liu(2019)}]{Zhang:2019ycw}
Zhang, X., \& Liu, S. 2019, arXiv:1903.02373

\bibitem[{{Zirakashvili} {et~al.}(2014){Zirakashvili}, {Aharonian}, {Yang},
  {O{\~n}a-Wilhelmi}, \& {Tuffs}}]{CasA_Zir&Ahar}
{Zirakashvili}, V.~N., {Aharonian}, F.~A., {Yang}, R., {O{\~n}a-Wilhelmi}, E.,
  \& {Tuffs}, R.~J. 2014, \apj, 785, 130

\end{thebibliography}
\end{document}